\documentclass[iop]{emulateapj}

\slugcomment{Accepted for Astrophysical Journal}

\shorttitle{The EBL from the far-UV to the far-IR}
\shortauthors{Driver et al.}

\begin{document}


\title{Extra-galactic background light measurements from the far-UV to
  the far-IR from deep ground and space-based galaxy counts}


\author{Simon P. Driver\altaffilmark{1}\altaffilmark{2}, Stephen K. Andrews, Luke J. Davies, Aaron S.G. Robotham, Angus H. Wright}
\affil{International Centre for Radio Astronomy Research (ICRAR), The University of Western Australia, M468, 35 Stirling Highway, Crawley, Australia, WA 6009}

\author{Rogier A. Windhorst, Seth Cohen, Kim Emig, Rolf A. Jansen}
\affil{School of Earth \& Space Exploration, Arizona State University, Tempe, AZ85287-1404, USA}
\and
\author{Loretta Dunne}
\affil{School of Physics and Astronomy, Cardiff University, Queen's Buildings, Cardiff, CF24 3AA, UK}
\affil{Institute for Astronomy, Royal Observatory, Blackford Hill, Edinburgh, EH9 3HJ, UK}


\altaffiltext{1}{(SUPA) School of Physics and Astronomy, University of St Andrews, North Haugh, St Andrews, KY16 9SS, UK}
\altaffiltext{2}{email: simon.driver@uwa.edu.au}


\begin{abstract}
We combine wide and deep galaxy number-count data from GAMA,
COSMOS/G10, {\it HST} ERS, {\it HST} UVUDF and various near-, mid- and
far- IR datasets from ESO, {\it Spitzer} and {\it Herschel}. The
combined data range from the far-UV (0.15$\mu$m) to far-IR
($500\mu$m), and in all cases the contribution to the integrated
galaxy light (IGL) of successively fainter galaxies converges. Using a
simple spline fit, we derive the IGL and the extrapolated-IGL in all
bands. We argue undetected low surface brightness galaxies and
intra-cluster/group light is modest, and that our extrapolated-IGL
measurements are an accurate representation of the extra-galactic
background light.  Our data agree with most earlier IGL estimates and
with direct measurements in the far-IR, but disagree strongly with
direct estimates in the optical. Close agreement between our results
and recent very high-energy experiments (H.E.S.S. and MAGIC), suggest
that there may be an additional foreground affecting the direct
estimates. The most likely culprit could be the adopted Zodiacal light
model. Finally we use a modified version of the two-component model
to integrate the EBL and obtain measurements of the Cosmic Optical
Background (COB) and Cosmic Infrared Background (CIB) of:
$24^{+4}_{-4}$ nW m$^{-2}$ sr$^{-1}$ and $26^{+5}_{-5}$ nW m$^{-2}$
sr$^{-1}$ respectively (48:52\%). Over the next decade, upcoming space
missions such as {\it Euclid} and {\it WFIRST}, have the capacity to
reduce the COB error to $<1\%$, at which point comparisons to the very
high energy data could have the potential to provide a direct
detection and measurement of the reionisation field.
\end{abstract}





\section{Introduction}
The extra-galactic background light, or EBL (\citealp{mcv59};
\citealp{par67a}; \citealp{par67b}; \citealp{hau01}; \citealp{lag05};
\citealp{kas06}; and \citealp{dwe13}), represents the flux received
today from a steradian of extragalactic sky. It includes all far-UV to
far-IR sources of photon production since the era of recombination,
and thereby encodes a record of the entire energy production history
of the Universe from $\sim$380,000yrs after the Big Bang to the
present day --- see \citet{wes87} and \citet{wes91} for an interesting
digression regarding the EBL's relation to Olber's Paradox. By convention,
the EBL is defined as the radiation received between 0.1$\mu$m to
1000$\mu$m (e.g., \citealp{fin10}, \citealp{dom11}; \citealp{dwe13};
\citealp{kha15}). This arises predominantly from star-light,
AGN-light, and dust reprocessed light --- with some minimal ($<15\%$)
contribution from direct dust heating due to accretion
(\citealp{ale05}; \citealp{jau11}). Photon production occurs not only
at far-UV to far-IR wavelengths, but across the entire
electro-magnetic spectrum (e.g., the cosmic x-ray background, see
\citealp{sha91}; and the cosmic radio background, see
\citealp{deo08}). However, based on integrated energy considerations
(i.e., $\int_{z=0}^{z=1050} \int^{\nu_2}_{\nu_1} \nu f_{\nu}\ \delta
\nu \ \delta z$), the cosmic emission is dominated, in terms of newly
minted photons, by the far-UV to far-IR range. Compared to the Cosmic
Microwave Background (CMB) the integrated-EBL is smaller by a factor
of $\sim20$, despite the $(1+z)^4$ diminution of the CMB photon
energies, but more than a factor of $\times100$ brighter than the
other backgrounds. Putting aside the CMB and the pre-recombination
Universe, the EBL is a product of the dominant astrophysical processes
which have taken place over the past 13~billion years, in terms of
energy redistribution (baryonic mass $\rightarrow$ photons). In
particular, because of the expansion, the precise shape of the
spectral energy distribution of the EBL depends on the cosmic
star-formation history, AGN activity history, and the evolution of
dust properties over cosmic time. It therefore represents rich
territory for comparison to galaxy formation and evolution models
(e.g., \citealp{dom11}; \citealp{som12}; \citealp{ino13}).

The EBL can be broken down into two roughly equal contributions from
the UV-optical-near-IR and the mid- to far- IR wavelength ranges: the
Cosmic Optical Background (COB; 0.1$\mu$m --- 8$\mu$m) and the Cosmic
Infrared Background (CIB $8\mu$m --- 1000$\mu$m;
\citealp{dwe98b}). Despite the different wavelength ranges, the COB
and CIB ultimately derive from the same origin: star-formation and
gravitational accretion onto super-massive black holes. The COB
represents the star- and AGN- light which directly escapes the host
system, and obsequiously pervades into the inter-galactic medium. The
CIB represents that component which is first attenuated by dust near
the radiation sources, and subsequently re-radiated in the mid- and
far-IR. The near equal balance between the energy of the integrated
COB and CIB is very much a testimony to the severe impact of dust
attenuating predominantly UV and optical photons, particularly given
the very modest amount of baryonic mass in the form of dust ($<1$\%
relative to stellar mass, see for example \citealp{dri08};
\citealp{dun11}).

Previous measurements of the EBL have come two flavours: direct
measurements (e.g., \citealp{pug96}; \citealp{fix98};
\citealp{dwe98a}; \citealp{hau98}; \citealp{lag99}; \citealp{dol06};
\citealp{ber02}; \citealp{ber07}; \citealp{cam01}; \citealp{mat05};
\citealp{mat11}) and integrated galaxy-counts (e.g., \citealp{mad00};
\citealp{hop10}; \citealp{xu05}; \citealp{tot01}; \citealp{dol06};
\citealp{kee10}; \citealp{ber11}; \citealp{bet12}). These two methods
should converge if the EBL is predominantly derived from galaxies
(including any AGN component), and if the photometric data used to
detect these galaxies are sufficiently deep. 

Until fairly recently, insufficient deep data existed to completely
resolve the EBL using galaxy number-counts, and direct measurements
appeared the more compelling constraint. However, with the advent of
space-based facilities ({\it GALEX}, {\it HST}, {\it Spitzer} and {\it
  Herschel}), and large ground-based facilities (VLT, Subaru), deep
field data have now been obtained across the entire far-UV to far-IR
range. The comparison of the direct estimates and integrated
number-counts are proving fertile ground for debate. In the CIB, the
direct estimates agree reasonably well with the integrated source
counts, which account for over 75\% of the directly measured CIB (see
\citealp{bet12} and \citealp{mag13}).  The remaining discrepancy can
be readily reconciled from extrapolations of the source counts, plus
some additional contribution from lensed systems (\citealp{war13}). In
the optical and near-IR the situation is less clear, with many direct
estimates being a factor of five or more greater than the integrated
galaxy counts (see for example the discussion on the near-IR
background excess in \citealp{kee10} or \citealp{mat15}), despite the
advent of very wide and deep data. Either the integrated source counts
are missing a significant quantity of the EBL in a diffuse component,
or the direct measures are over-estimated (i.e., the backgrounds are
under-estimated).

Recently a third pathway to the EBL has opened up, via the indirect
attenuation of TeV photons emanating from Blazars, as observed with
Very High Energy (VHE) experiments (e.g., the High Energy Stereoscopic
System, H.E.S.S., and the Major Atmospheric Gamma Imaging Cherenkov
telescope, MAGIC). Here the TeV flux from a distant Blazar, believed
to extrude a well behaved power-law spectrum, interacts with the
intervening EBL photon-field. Preferential interactions between TeV
and micron photons create electron-positron pairs, thereby removing
power from the received TeV spectrum over a characteristic wavelength
range. The proof of concept was demonstrated by \citet{aha06} and
comprehensive measurements have recently been made by both the
H.E.S.S. Consortium (\citealp{abr13}) and the MAGIC team
(\citealp{ahn16}). These two independent measurements very much favour
the low-EBL values. However, uncertainty remains as to the strength,
and hence, impact of the inter-galactic magnetic field, the intrinsic
Blazar spectrum shape, and the role of secondary PeV
cascades. Moreover, the two VHE studies mentioned above require a
pre-defined EBL model, and use the shape of the received TeV signal
compared to the assumed intrinsic spectrum, to provide a normalisation
point. Hence spectral information is essentially lost. Trickier to
determine but more powerful is the potential for the VHE data to
constrain both the normalisation {\it and} the shape of the EBL
spectrum. A first effort was recently made by \cite{bit15} who again
found results consistent with the low-EBL value, but perhaps more
crucially provided independent confirmation of the overall shape
across the far-UV to far-IR wavelength range. Nevertheless, caveats
remain as to the true intrinsic Blazar spectral slope in the TeV
range, contaminating cascades from PeV photons, the strength of any
intervening magnetic fields, and in some cases the actual redshift of
the Blazars studied.

Here we aim to provide the first complete set of integrated-galaxy
light measurements based on a combination of panchromatic wide and
deep number/source-count data from a variety of surveys which
collectively span the entire EBL wavelength range (0.1$\mu$m ---
1000$\mu$m). Our approach varies from previous studies, in that we
abandon the concept of modeling the data with a galaxy-count (backward
propagation) model. Instead, as the number-count data is bounded in all
bands --- in terms of the contribution to the integrated luminosity
density --- we elect to fit a simple spline to the luminosity density
data.

In Section~2, we present the adopted number-count data, the trimming
required before fitting, and estimates of the cosmic (sample) variance
associated with each dataset. In Section~3, we describe our fitting
process and the associated error analysis. We compare our measurements
to previous estimates in Section 4, including direct and VHE
constraints, before exploring possible sources of missing light. We
finish by using the EBL model of Driver \& Andrews (see
\citealp{dri13} and \citealp{and16b}) as an appropriate fitting
function to derive the total integrated energy of the COB and CIB.

All magnitudes are reported in the AB system, and where relevant we
have used a cosmology with $\Omega_{\Lambda}=0.7, \Omega_{\rm
  Matter}=0.3$, and, $H_o=70$km/s/Mpc.

\section{Number-count data}
This work has been motivated by the recent availability of a number of
panchromatic data sets which extend from the far-UV to the far-IR. In
particular: Galaxy And Mass Assembly (GAMA) (\citealp{dri11};
\citealp{dri16a}; \citealp{lis15}; \citealp{hop13}); COSMOS/G10
(\citealp{dav15}; \citealp{and16a}); {\it Hubble Space Telescope}
Early Release Science (ERS; \citealp{win11}); and the {\it Hubble Space
  Telescope} Ultra-Violet Ultra-Deep Field (UVUDF; \citealp{tep13},
\citealp{raf15}). For each of these datasets great care has been taken
to produce consistent and high quality photometry across a broad
wavelength range. We have been responsible for the production of the
first three catalogues, with the fourth recently made publicly
available. We describe the various datasets in more detail below.

\subsection{Galaxy And Mass Assembly (GAMA)}
GAMA represents a survey of five sky regions covering 230~$\deg^2$
\citep{dri11}, with spectroscopic data to $r=19.8$mag obtained from
the Anglo-Australian Telescope \citep{lis15}.  Complementary
panchromatic imaging data comes from observations via {\it GALEX}
\citep{lis15}, SDSS \citep{hil11}, VISTA \citep{dri16a}, {\it WISE}
\citep{clu14}, and {\it Herschel} (\citealp{eal10},
\citealp{val16}). The GAMA Panchromatic Data Release imaging
\citep{dri16a} was made publicly available in August 2015
(\url{http://gama-psi.icrar.org}). The 180~$\deg^{2}$ which makes up
the three GAMA equatorial regions (G09, G12 \& G15), has since been
processed with the custom built panchromatic analysis code {\sc
  lambdar} \citep{wri16}, to derive matched-aperture and PSF-convolved
photometry across all bands. The Wright-catalogue is $r$-band selected
using SExtractor \citep{ber96}. The {\sc SExtractor}-defined apertures
are convolved with the appropriate point-spread function for each
facility, and used to derive consistent flux measurements in all other
bands.  Care is taken to manage overlapping objects and flux share
appropriately (see \citealp{wri16} for full details). As part of the
quality control process, all bright and all oversized apertures (for
their magnitude), were visually inspected and corrected if
necessary. The Wright-catalogue then uses these apertures to provide
fluxes in the following bands: FUV/NUV, $ugri$, $ZYJHK$,
$IRAC$-$1/2/3/4$, PACS~100/160, SPIRE~250/350/500.  Number-counts are
generated by binning the data within 0.5mag intervals and scaling for
the area covered. Random errors are derived assuming Poisson statistics
(i.e., $\sqrt{n}$) with cosmic variance errors included in the
analysis as described in Sections 2.5 and 3.3.4.  Note that in
generating number counts, the $r$-band selection will lead to a
gradual flattening of the counts in other bands, particularly those
bands furthest in wavelength from $r$, i.e., a colour bias. The very
simple strategy we adopt here is to identify the point at which the
GAMA Wright-catalogue counts diverge from the deeper datasets and
truncate our counts 0.5~mag brightwards of this limit. See
\citet{wri16} for full details of the GAMA analysis including aperture
verification.

\subsection{COSMOS/G10}
COSMOS/G10 represents a complete reanalysis of the available
spectroscopic and imaging data to GAMA standards, in a 1~sq.~degree
region of the {\it HST} COSMOS field \citep{sco07}, dubbed G10 (GAMA
$10^h$; see \citealp{dav15}, \citealp{and16a} for full details). The
primary source detection catalogue uses SExtractor applied to deep
Subaru $i$-band data, following trial and error optimisation of the
SExtractor detection parameters. All anomalous apertures were manually
inspected and repaired, as needed. As for GAMA, the {\sc lambdar}
software was used to generate matched aperture photometry from the
far-UV to far-IR. The photometry for the COSMOS/G10 region spans 38
wavebands, and combines data from {\it GALEX} \citep{zam07}, CFHT
\citep{cap07}, Subaru \citep{tan07}, VISTA \citep{mcc12}, {\it
  Spitzer} (\citealp{san07} and \citealp{fra09}) and {\it Herschel}
(\citealp{lut11}, \citealp{oli12}, \citealp{lev10}, \citealp{vie13},
\citealp{smi12} and \citealp{wan14}). As for GAMA the number-counts
are derived in 0.5mag bins yielding counts in: FUV/NUV, u, griz, YJHK,
$IRAC 1/2/3/4$, MIPS~24/70, PACS~100/160, SPIRE~250/350/500. The
COSMOS/G10 catalogue is $i$-band selected, and hence a gradual decline
in the number counts will occur in filters other than $i$, as either
very red or blue galaxies are preferentially lost. In particular, a
cascading flux cut was implemented as the optical priors were advanced
into the far-IR to minimise erroneous measurements.  As for GAMA we
use the departure of the COSMOS/G10 counts from the available deeper
data to identify the magnitude limits at which the counts become
incomplete. See \citet{and16a} for full details of the COSMOS/G10
analysis including data access. 

\subsection{Hubble Space Telescope Early Release Science ({\it HST} ERS)}
The {\it HST} ERS dataset \citep{win11} represents one of the first
fields obtained using {\it HST}'s WFC3, and built upon earlier optical
ACS imaging of the GOODS South field. The analysis details are
provided in Windhorst et al (2011), and counts are derived in 11
bands: F225W, F275W, F306W, F435W, F775W, F850LP, F098M, F105W, F125W,
F140W, and F160W covering 40-50 arcmin$^2$ to AB
$26.3$---$27.5$~mag. For the {\it HST} ERS, data catalogues were
derived independently in each band, hence avoiding any color
bias. {\it Spitzer} and {\it Herschel} data exists for the {\it HST}
ERS field, but is currently not part of this analysis due to its much
coarser beam and the resulting faint-end confusion. In due course, we
expect to reprocess the ERS data using {\sc lambdar}. In addition to
the {\it HST} ERS field the ASU team have also measured deep counts in
various bands in the UDF and XDF fields using an identical process as
that described for the {\it HST} ERS data, but including a correction
for incompleteness by inserting false galaxy images into real data and
assessing the fraction recovered as a function of magnitude. See
\cite{win11} for full details of the {\it HST} ERS and deep field
analysis.

\subsection{Hubble Space Telescope ultra-violet ultra-deep field ({\it HST} UVUDF)}
The {\it HST} UVUDF dataset (\citealp{tep13}) represents a major
effort to bring together data on the HUDF field spanning a broad
wavelength range and consisting of 11 bands from the near-UV to the
near-IR (see \citealp{raf15}).  The UVUDF team also use an aperture
matched point-spread function corrected method to derive the
photometry. The area of coverage varies from 7.3 to 12.8 arcmin$^2$
across the bands.  The initial detection image is derived from the
combination of 4 optical and 4 near-IR bands (weighted by the inverse
variance of each image on a pixel-by-pixel basis). As the near-IR data
is variable in depth, this does produce a catalogue with a slightly
variable detection wavelength. However, the basic strategy of object
detection based on multi-filter stacked data combined with pixel
weighting, should mitigate most of the color selection bias in the
count data. The Rafelski et al.~catalogue is publicly
available\footnote{\url{http://uvudf.ipac.caltech.edu/}} and counts
are derived by binning the data in 0.5mag intervals and scaling for
area. The {\it HST} UVUDF dataset contains counts in the following
bands: F225W, F275W, F336W, F435W, F606W, F775W, F850LP, F105W, F125W,
F140W, and F160W.  See \citet{raf15} for full details of the UVUDF
data analysis.

\subsection{Additional datasets \& cosmic variance (CV)}
In addition to the four primary catalogues mentioned above, we include
a number of band specific surveys to extend the range of our number
count analysis into the far-UV, mid-IR and far-IR. These
  datasets, along with the four primary datasets, are summarised in
  Table~\ref{tab:datasets}. All datasets will be susceptible to cosmic
  variance particularly at the fainter ends of the {\it HST} datasets,
  where the volumes probed will be very small. For each dataset we
  therefore derive a cosmic (sample) variance error; listed in Col.~5
  of Table~\ref{tab:datasets} using equation 4 from \citet{dri10}.
  These values are based on the quoted field-of-view, the number of
  independent fields, and an assumed redshift distribution and the
  adopted values are also shown in Table.~\ref{tab:datasets}.  We
  refer the curious reader to our online calculator: {\url
    http://cosmocalc.icrar.org} (see Survey Design tab).

\begin{table*}
\caption{A summary of datasets used in this analysis and in particular
  the cosmic variance errors estimated via Eqn~4 from \citet{dri10},
  see also http://cosmocalc.icrar.org for our online
  calculator. \label{tab:datasets}}
\begin{tabular}{cp{4.75cm}ccrl} \\ \hline
Filter & Description & $\Omega$ & z & CV & Reference \\
       &             & ($\deg^2$) & range & (\%) & \\ \hline
All & GAMA 21 band panchromatic data release & $3 \times 60$ & 0.05---0.30 & 5\% & \citet{wri16} \\
All & COSMOS/G10 38 band panchromatic data release & 1 & 0.2---0.8 & 11\% & \citet{and16a} \\
UV-NIR & {\it HST} WFC3 Early Release Science (Panchromatic Counts) & 0.0125 & 0.5---2.0 & 17\% & \citet{win11} \\
UV-NIR & {\it HST} Ultra-violet through NIR observations of the Hubble UDF & 0.0036 & 0.5---2.0 & 21\% & \citet{raf15} \\
FUV & {\it HST} ACS Solar Blind Channel observations of HDF-N(GOODS-S), HUDF, GOODS-N & $3 \times \sim0.0015$& 0.5---2.0 & 14\% & \citet{voy11}\\
NUV & STIS observations in the HDF-N, HDF-S, and the HDF-parallel & $3 \times \sim0.00012$ & 0.5---2.0 & 21\% & \citet{gar00} \\
$u$ & near-UV(360nm) observations in the Q0933+28 field with the Large Binocular Camera & 0.4 & 0.5---2.0 & 8\% & \citet{gra09} \\
$K$ & The Hawk-I UDS and GOODS Survey (HUGS) & 0.016, {\bf 0.078} & 0.5---2.0 & 11\% & \citet{fon14} \\
$IRAC$-$1/2$ & S-CANDELS: The Spitzer-Cosmos Assembly Near Infrared Deep Extragalactic Survey: 3.6 and 4.5 $\mu$m {\it Spitzer} IRAC & $5 \times \sim0.032$ & 0.5---2.0 & 6\% & \citet{ash15} \\
$IRAC$-$4$(8$\mu$m) & {\it Spitzer} IRAC Band 4 data of the EGS field & 0.38 & 0.5---2.0 & 8\% & \citet{bar08}\\
$24\mu$m & {\it Spitzer} MIPS 24$\mu$m observations of Marano, CDF-S, EGS, Bo\"otes and ELAIS fields & 0.36,0.58,0.41,{\bf 9.0},0.036 & 0.5---2.0 & 3\% & \citet{pap04} \\ 
$24\mu$m & {\it Spitzer} MIPS 24$\mu$m data in FIDEL, COSMOS and SWIRE fields & {\bf 53.6} (9 fields) & 0.5---2.0 & 2\% & \citet{bet10} \\
$70/160 \mu$m & {\it Spitzer} MIPS 70 \& 160 $\mu$m data & {\bf 45.4} (9 fields) & 0.5---2.0 & 2\% & \citet{bet10} \\
$70 \mu$m & {\it Herschel} PACS 70 $\mu$m data & {\bf 42.9} (9 fields) & 0.5---2.0 & 2\% & \citet{bet10} \\
$100/160 \mu$m & {\it Herschel} The PACS Evolutionary Probe Survey (100 \& 160 $\mu$m data) of the GOODS-S, GOODS-N, Lockman Hole and COSMOS areas & 0.083,0.083,0.18,{\bf 2.04} & 0.5---2.0 & 5\% & \citet{ber11} \\
$100/160 \mu$m & {\it Herschel} PACS 100 \& 160 $\mu$m data from PEP and GOODS-Herschel campaigns & $2 \times \sim0.052$ & 0.5---2.0 & 9\% & \citet{mag13} \\
$250/350/500 \mu$m & {\it Herschel} (HerMES) SPIRE 250, 350 \& 500 $\mu$m observations in the COSMOS and GOODS-N regions & 0.083,{\bf 2.04} & 0.5---2.0 & 5\% & \citet{bet12} \\ 
$250/350/500 \mu$m & {\it Herschel} (H-Atlas) SPIRE 250, 350 \& 500 $\mu$m observations in the equatorial GAMA fields & $3 \times 54$ & 0.05---2.0 & 2\% & \citet{val16} \\ \hline
\end{tabular}
\end{table*}

\begin{figure}[h]

\epsscale{1.0}

\plotone{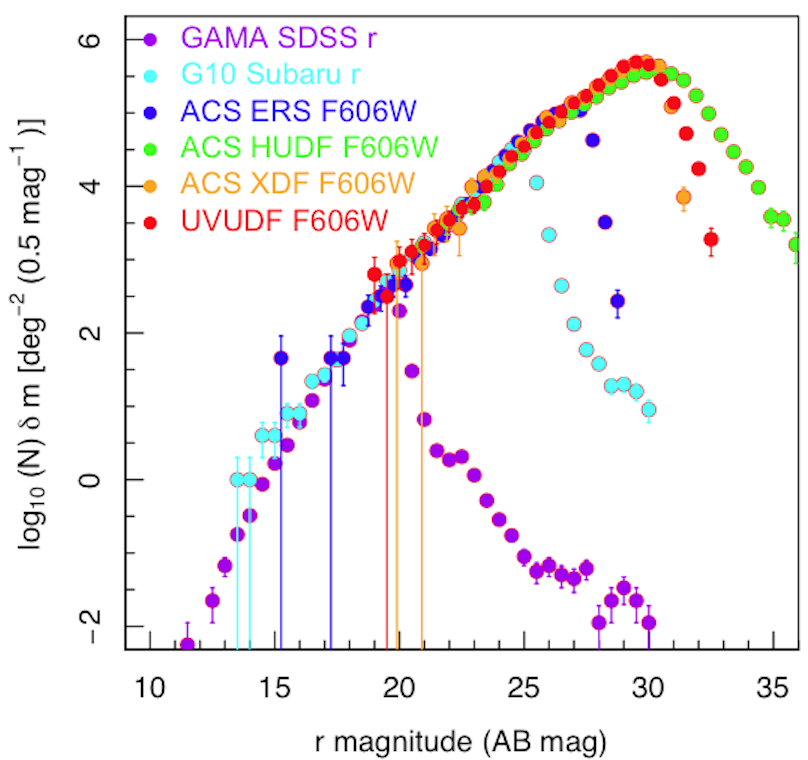}

\caption{$r$-band number counts produced by simply combining the
  catalogues as provided.  Each dataset shows a clear down-turn of the
  number counts at faint magnitudes, which is an artifact of the flux
  and color limit of that dataset. To correct for this, we truncate
  each dataset 0.5~mag brightwards of its turn-down point. In some
  cases we also truncate bright data, when the errorbars become
  excessive, and would have no impact on our fitting whether included
  or not. \label{fig:counts}}
\end{figure}

\subsection{Merging datasets}
\begin{figure*}

\epsscale{1.0}

\plotone{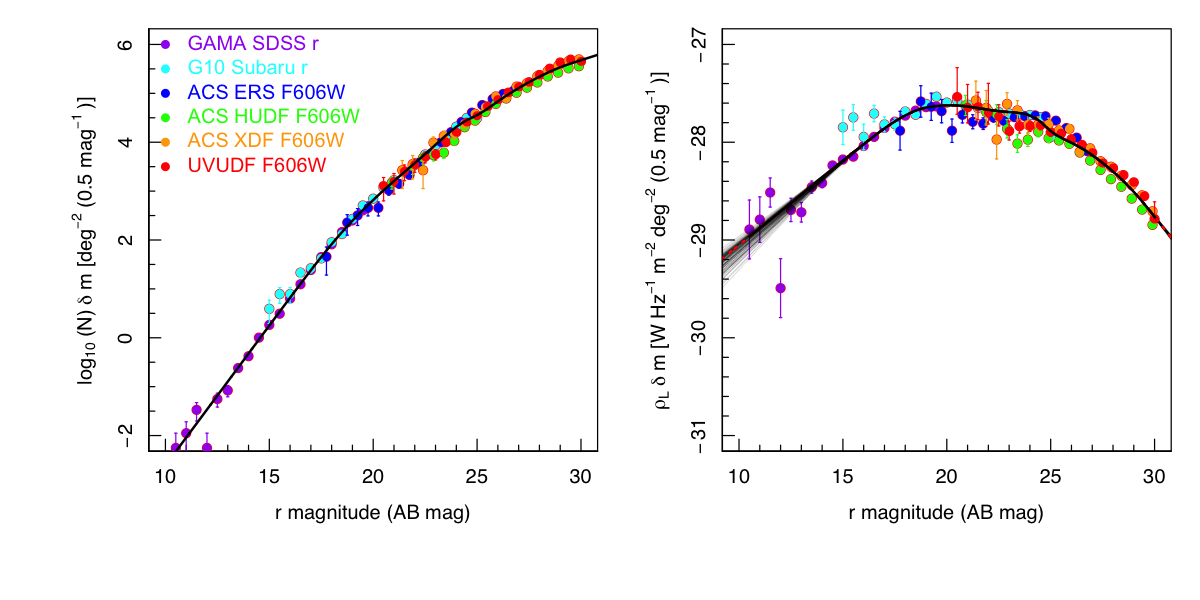}

\vspace{-1.0cm}

\plotone{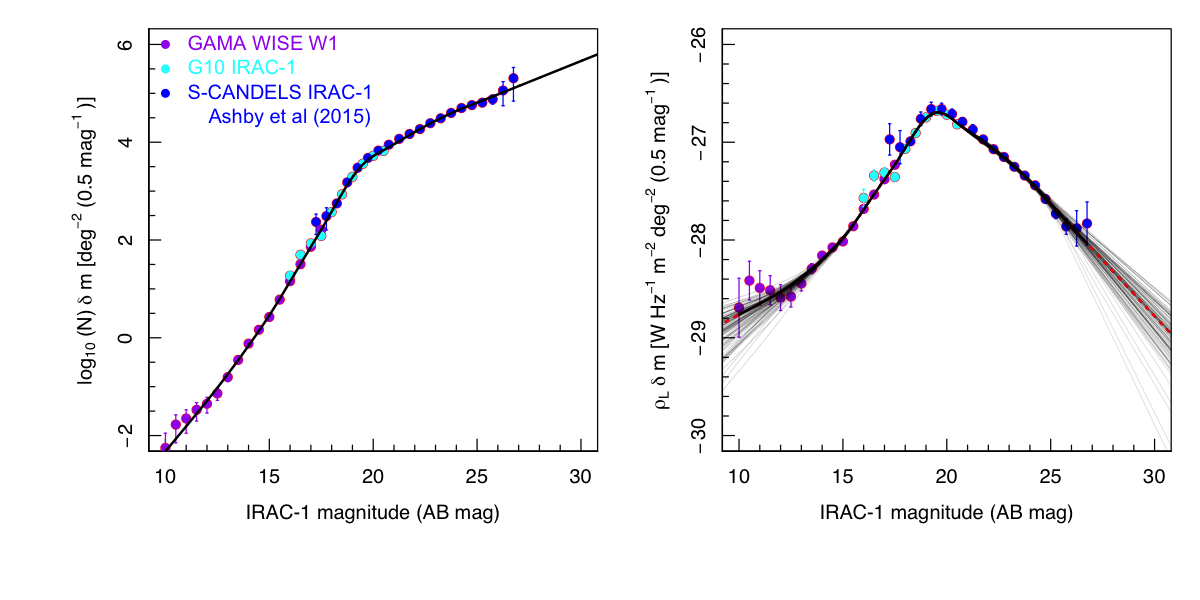}

\vspace{-1.0cm}

\plotone{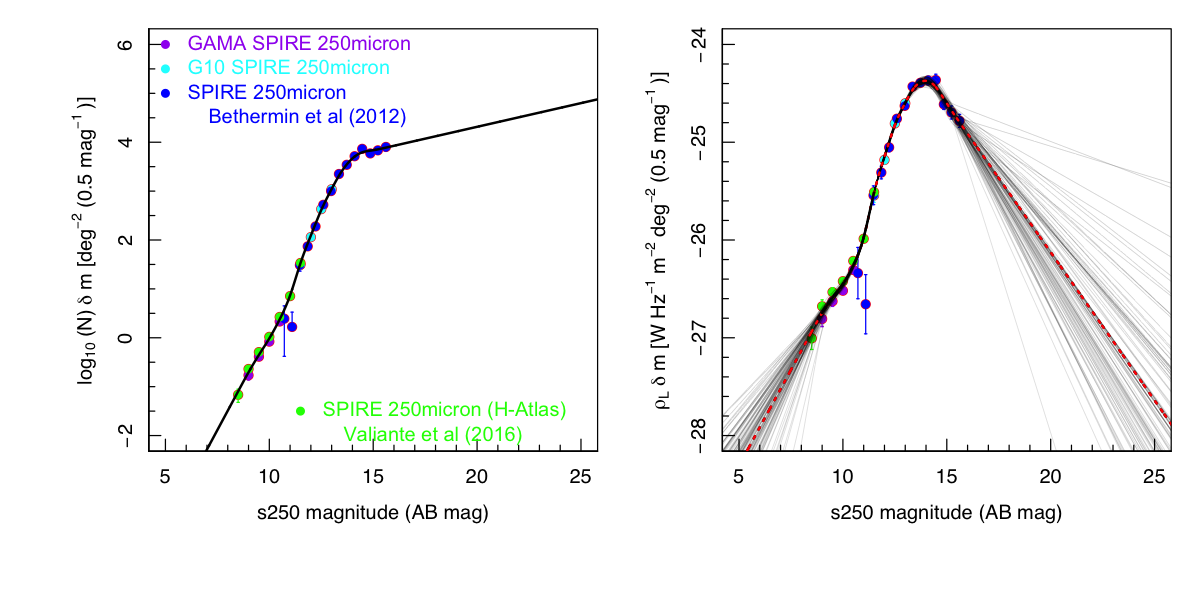}

\vspace{-1.0cm}

\caption{(left) Galaxy number counts in the specified band, where the
  black curve depicts the 10-point spline fit to the luminosity
  density data. (right) The contribution of each magnitude interval to
  the luminosity density. Where the spline-fit is extrapolated this is
  shown as a red dashed line. Also shown are 101 faint grey lines
  indicating identical fits to data randomly perturbed within their
  errorbars.
\label{fig:ebl}}

\end{figure*}

Fig.~\ref{fig:counts} shows the combined galaxy number counts in the
$r$ band from six datasets of varying areas and depths. The figure
highlights the distinct limits of each dataset reflected by the abrupt
turn-downs. Care must therefore be taken to truncate each dataset at
an appropriate magnitude limit. For the GAMA and COSMOS/G10 datasets
we identify the turn-down as the point at which the data become
inconsistent with the deeper datasets. This is because the colour bias
introduces a shallow rather than abrupt turn-down. For all other
datasets we identify the point at which the counts at faint
magnitudes fall abruptly, and then truncate 0.5~mag
brightwards. Truncating in this way results in a fairly seamless
distribution (see Fig.~\ref{fig:ebl}, top left panel).

Our final number-count distributions, for three arbitrarily selected
bands ($r$, $IRAC$-$1$ and SPIRE~$250\mu$m), are shown in Fig.~\ref{fig:ebl}
(left panels). Following the trimming process, the datasets shown
overlap extremely well.

Finally, the reader will notice that data contributing to the number
counts in some bands are determined through non-identical
filters. Perfect color corrections to a single bandpass would require
individual SED fitting, which itself is imprecise, and the corrections
of the mean of the data will in all cases be less than $\pm
0.05$~mag. Given the very good agreement between our counts, despite
slight filter discrepancies, we elect to assume that these offsets do
not significantly affect our derived results, and to instead fold in
an additional 0.05~mag systematic error into our EBL error analysis
(see Section~\ref{Sec:photom}).

\section{The far-UV --- far-IR extra-galactic background light (EBL)}
To derive an extrapolated integrated galaxy count (eIGL) measurement
from galaxy count data, one typically constructs a galaxy number count
model tailored to match the data. One can then integrate the
luminosity weighted model to either the limit of the data (providing a
lower limit), or to the limit of the model (an extrapolated
measurement). In practice, these measures are referred to as
integrated galaxy light measurements (IGL), and in the absence of
other significant sources of radiation, should equate to the true
EBL. Here, we deviate from this path in two ways: Firstly, we elect to
simply fit a 10-point spline to the available data; and secondly, we
directly fit to the luminosity-weighted data rather than the number
counts. These two departures are intended to provide a more robust
measurement as the spline should map the nuances of the data
perfectly, while number-count models are inevitably imperfect. In the
event that the distributions are well bounded by the data in terms of
their contribution to the IGL, the non-physical nature of the
extrapolation is not particularly significant.

\subsection{Spline fitting}
We derive our luminosity density values via spline fitting, using
  the $R$\footnote{\cite{r15}. R: A language and environment for
    statistical computing. R Foundation for Statistical Computing,
    Vienna, Austria. \url{https://www.R-project.org/}} {\sc
    smooth.spline} routine with 10 degrees of freedom (spline
  points). In fitting the spline the data points are weighted
  inversely proportional with $\sigma^2$. To derive our IGL estimates
  we use the spline fit to populate a differential luminosity density
  distribution from AB=$-$100 to AB=100 mag in 0.01mag intervals and
  then sum the predicted values (dividing by the bin width). In all
  cases, flux outside the summing range is negligible.  Where datasets
  overlap in a particular magnitude interval, our spline fits, will be
  driven by the survey with the largest area coverage --- as the
  fitting process is error weighted ($1/\sigma^2$). Hence, the
  significance of the fit will progress from data drawn from 180
  deg$^2$ for GAMA at the brightest end, through 1 deg$^2$ for the
  COSMOS/G10 region to 40 arcmin$^2$ for the {\it HST} ERS data and 10
  arcmin$^2$ for the deepest {\it HST} UDF data. This progression of
  area and depth highlights the importance of combining both wide and
  deep datasets in this way. Number-count data for the FUV/NUV, $ugi$,
  $ZYJHK$, {\it IRAC}~124, MIPS24, {\it Herschel} PACS~70/100/160 and
  SPIRE~250/350/500 bands are shown in Figs. \ref{fig:ebl2} ---
  \ref{fig:ebl3}. In all cases, the number-count data are consistent
  across the datasets within the quoted errors following the above
  trimming process.

\subsection{Measurements of the EBL}
Fig.~\ref{fig:ebl} (right panels) shows the contribution within each
magnitude interval to the luminosity density (data points as
indicated) for the $r$, $IRAC$-$1$, and SPIRE~250$\mu$m
bands. Overlaid is the best-fit spline model (black curve).  Beyond
the data range, the extrapolation of the spline-fit are shown as
red-dashed lines. Also shown as grey lines, are spline-fits to
perturbations of the data as described in
Section~\ref{sec:errors}. Note that the units we adopt for the far-IR
data will be unfamiliar to the far-IR community used to working in
Euclidean normalised source counts in intervals of Jansky. Here, for
consistency, we have elected to process and show all data in the
traditional optical units of AB magnitude
intervals. Figs~\ref{fig:ebl2} to~\ref{fig:ebl3} (right panels) show
the luminosity density fits in the remaining 18 bands. In all cases
the data is bounded (right side panel), i.e., the contribution to the
luminosity density rises to a peak and then decreases with increasing
magnitude (decreasing flux). This implies that the dominant
contribution to the IGL is resolved, and that adopting a
spline-fitting approach rather than a galaxy number-count model
approach is reasonable. The only possible exception is the $IRAC$-$4$
data, where the peak is only just bound (see Fig.\ref{fig:w3}, lower
panel). 

\begin{table*}
\caption{Measurements of the eIGL from integrating spline fits to the
  luminosity weighted number counts (Col.~3) and from the median of
  our 10,001 Monte-Carlo realizations (Col.~4). Lower limits to the
  eIGL from integrating within the data range only (Col.~5), and
  errors (associated with zero-point uncertainty, fitting methodology,
  and from Monte-Carlo realizations of the random or CV errors
  (Cols.~6,~7 \&~8 respectively), all at the wavelengths indicated by
  the bandpass or pivot wavelength indicated in Col.~1 or~2
  respectively.
\label{tab:ebl}}
\begin{center}
\begin{tabular}{ccccccccc} \hline \hline
Filter & Pivot      & Extrapolated & Extrapolated & IGL Lower & Zeropoint & Fitting & Poisson & CV \\
Name & Wavelength & IGL (best-fit) & IGL (median)   & limit & error      & error  & error & error \\  
& ($\mu$m)  & \multicolumn{6}{c}{(nW m$^{-2}$ sr$^{-1}$)} \\ \hline 
Col.~1 & Col.~2 & Col.~3 & Col.~4 & Col.~5 & Col.~6 & Col.~7 & Col.~8  & Col.~9\\ \hline
FUV       &   0.153  &    1.45  &    1.45  &    1.36  & $\pm   0.07 $ & $\pm   0.00 $ & $\pm   0.04 $ & $\pm   0.16 $\\  
NUV       &   0.225  &    3.15  &    3.14  &    2.86  & $\pm   0.15 $ & $\pm   0.02 $ & $\pm   0.05 $ & $\pm   0.45 $\\  
$u$       &   0.356  &    4.03  &    4.01  &    3.41  & $\pm   0.19 $ & $\pm   0.04 $ & $\pm   0.09 $ & $\pm   0.46 $\\  
$g$       &   0.470  &    5.36  &    5.34  &    5.05  & $\pm   0.25 $ & $\pm   0.04 $ & $\pm   0.05 $ & $\pm   0.59 $\\  
$r$       &   0.618  &    7.47  &    7.45  &    7.29  & $\pm   0.34 $ & $\pm   0.05 $ & $\pm   0.04 $ & $\pm   0.69 $\\  
$i$       &   0.749  &    9.55  &    9.52  &    9.35  & $\pm   0.44 $ & $\pm   0.00 $ & $\pm   0.05 $ & $\pm   0.92 $\\  
$z$       &   0.895  &   10.15  &   10.13  &    9.98  & $\pm   0.47 $ & $\pm   0.03 $ & $\pm   0.05 $ & $\pm   0.96 $\\  
$Y$       &   1.021  &   10.44  &   10.41  &   10.23  & $\pm   0.48 $ & $\pm   0.00 $ & $\pm   0.07 $ & $\pm   1.05 $\\  
$J$       &   1.252  &   10.38  &   10.35  &   10.22  & $\pm   0.48 $ & $\pm   0.00 $ & $\pm   0.05 $ & $\pm   0.99 $\\  
$H$       &   1.643  &   10.12  &   10.10  &    9.99  & $\pm   0.47 $ & $\pm   0.01 $ & $\pm   0.06 $ & $\pm   1.01 $\\  
$K$       &   2.150  &    8.72  &    8.71  &    8.57  & $\pm   0.40 $ & $\pm   0.02 $ & $\pm   0.04 $ & $\pm   0.76 $\\  
$IRAC$-$1$  &   3.544  &    5.17  &    5.15  &    5.03  & $\pm   0.24 $ & $\pm   0.03 $ & $\pm   0.06 $ & $\pm   0.43 $\\  
$IRAC$-$2$  &   4.487  &    3.60  &    3.59  &    3.47  & $\pm   0.17 $ & $\pm   0.02 $ & $\pm   0.05 $ & $\pm   0.28 $\\  
$IRAC$-$4$  &   7.841  &    2.45  &    2.45  &    1.49  & $\pm   0.11 $ & $\pm   0.77 $ & $\pm   0.15 $ & $\pm   0.08 $\\  
MIPS24    &  23.675  &    3.01  &    3.00  &    2.47  & $\pm   0.14 $ & $\pm   0.05 $ & $\pm   0.06 $ & $\pm   0.07 $\\  
MIPS70    &  70.890  &    6.90  &    6.98  &    5.68  & $\pm   0.32 $ & $\pm   0.07 $ & $\pm   0.79 $ & $\pm   0.18 $\\  
PACS100   & 101.000  &   10.22  &   10.29  &    8.94  & $\pm   0.47 $ & $\pm   0.10 $ & $\pm   0.56 $ & $\pm   0.88 $\\  
PACS160   & 161.000  &   16.47  &   16.46  &   10.85  & $\pm   0.76 $ & $\pm   0.81 $ & $\pm   2.99 $ & $\pm   1.57 $\\  
PACS160$^\dagger$   & 161.000  &   13.14  &   9.17  &    8.93  & $\pm   0.61 $ & $\pm   0.17 $ & $\pm   1.32 $ & $\pm   0.72 $\\  
SPIRE250  & 249.000  &   10.00  &   10.04  &    8.18  & $\pm   0.46 $ & $\pm   0.18 $ & $\pm   0.87 $ & $\pm   0.59 $\\  
SPIRE350  & 357.000  &    5.83  &    5.87  &    4.66  & $\pm   0.27 $ & $\pm   0.24 $ & $\pm   1.04 $ & $\pm   0.32 $\\  
SPIRE500  & 504.000  &    2.46  &    2.48  &    1.71  & $\pm   0.11 $ & $\pm   0.03 $ & $\pm   2.54 $ & $\pm   0.13 $\\ \hline 
\end{tabular}
\end{center}
$^\dagger$ re-fitted excluding the very faint number-count data of
\citet{mag13} where the completeness corrections exceeds $\times 1.2$
\end{table*}

\begin{figure*}

\plotone{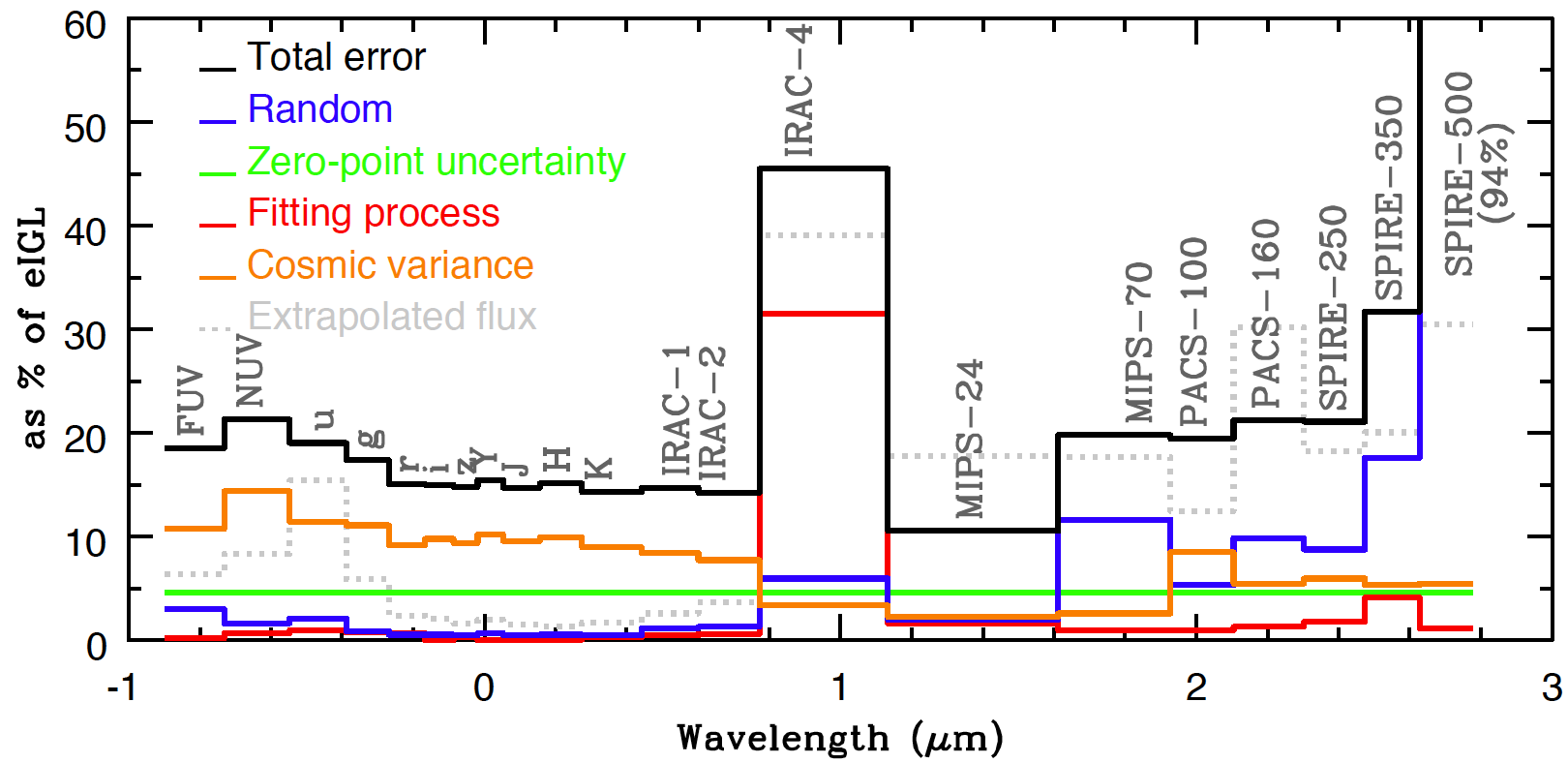}

\caption{The contribution of each of the individual errors as a
  percentage of the eIGL measurement as a function of wavelength (as
  indicated). The back line shows the total error which is mostly
  dominated by cosmic variance in the optical,near and mid-IR bands
  but by random errors in the far-IR bands. \label{fig:errors}}
\end{figure*}

In Table~\ref{tab:ebl} we present three distinct measurements. The
first (Col.~3) is our best-fit eIGL values based on a spline fit to
the data. We also present the median value from 10,001 Monte-Carlo
realizations (Col.~4) as discussed in Section~\ref{sec:errors}. In all
cases, the best-fit and median estimates agree extremely closely as
one would expect.  In Col.~5 we present the measurement of the IGL,
but confine ourselves to the range covered by the data, i.e., no
extrapolation. This will naturally provide a lower limit, and a
comparison between the values in Col.~3 and Col.~5 provides some
indication of where the extrapolation is important for measuring the
eIGL. In most cases (see Fig.~\ref{fig:errors}, grey dotted lines),
comparisons between the lower-limit and extrapolated values suggest
that $<5-10\%$ of the COB measurements derive from these
extrapolations, and typically $<20-30$\% for the CIB. Although the
spline fit is not physically motivated, the figures show that the fits
behave sensibly, and that the extrapolations project linearly beyond
the range of the data points.  However, as the underlying number-count
data is generally flattening (because of the diminishing volume for
higher-z systems due to the cosmological expansion), this does lead to
the possibility of a small over-estimate in our eIGL measurements,
albeit well within the quoted errors. Hence, the most cautious way to
interpret our analysis would be to adopt the range which extends from
the lower limit from the non-extrapolated values (minus the error) to
the eIGL (plus the error).


\begin{table*}
\caption{The compendium of galaxy number counts in 21 bands assembled
  from various sources and contained in one machine readable file. A
  sample of the first and last three lines of the data file are shown
  here. Cols.1\&2 indicate the facility and filter from which the data
  are derived. Col.3 the magnitude bin centre, Col.4 the number counts
  within that bin, and Col.5 the error as provided. Col.6 refers to
  the dataset number for that filter. Col.7 the cosmic variance as
  shown in Table~\ref{tab:datasets} and Col.8 the literature reference
  for the data.
\label{tab:counts}}
\begin{center}
\begin{tabular}{cccccccc} \hline \hline
Facility & Filter & Mag. bin    & N(m) & $\Delta$ N(m) & Seq. & Cos. Var. & Reference \\
Name     & Name   & center (mag) & 0.5mag$^{-1} \deg^{-2}$ & 0.5mag$^{-1} \deg^{-2}$ & No. & (\%) & \\ \hline
GALEX	&FUV	&14.0	&0.01331	&0.00941	&1	&5	&Wright et al. (2016) \\
GALEX	&FUV	&14.5	&0.01331	&0.00941	&1	&5	&Wright et al. (2016) \\
GALEX	&FUV	&15.0	&0.01996	&0.01152	&1	&5	&Wright et al. (2016) \\
... \\
Herschel	&SPIRE500	&14.8586	&4107.05	&617.418	&3	&5	&B\'ethermin et al. (2012) \\
Herschel	&SPIRE500	&15.2244	&4898.7	        &735.368	&3	&5	&B\'ethermin et al. (2012) \\
Herschel	&SPIRE500	&15.612	        &5556.47	&1385.51	&3	&5	&B\'ethermin et al. (2012) \\ \hline
\end{tabular}
\end{center}
Notes: Table~\ref{tab:counts} is published in its entirety in a
machine readable format. A portion is shown here for guidance
regarding its form and content.
\end{table*}

Table~\ref{tab:counts} provides an extract of our trimmed data listing
our compiled number counts and associated errors, as used for our
spline fitting for each wave-band. This data file in machine readable
format from the ApJ online edition.

\subsection{Appropriate error estimation}
There are a number of possible sources of error; in particular we
consider those arriving from a systematic photometry error (zero-point
shift, or background over/under estimation), the fitting process,
those implied from the errors in the count data, and the possible
impact of cosmic variance. We explore each of these in turn.

\subsubsection{Photometric error \label{Sec:photom}}
In most cases where we have multiple datasets we see that the counts
agree within 0.05~mag, despite the potential for filter offsets due to
small bandpass discrepancies. Generally, errors in absolute
zero-points, particularly in {\it HST} data are expected to be
$<0.01$---$0.02$~mag. However the process of sky-subtraction, object
detection, and photometric measurement can lead to significant {\it
  systematic} variations. The easiest way to quantify the impact is to
{\it systematically} shift all datasets by $\pm 0.05$~mag and
re-derive our measurements. The value of $\pm 0.05$~mag comes from the
amount required to align the various deep datasets, and is taken here
to represent the systematic uncertainty in the entire photometric
extraction process.  This level of uncertainty should be considered
conservative --- surveys such as GAMA, for example typically quote
errors of $\pm 0.03$ in photometric measurements --- but high-$z$
galaxies are often asymmetric, and their photometry distorted by
ambiguous deblends. Hence, a larger assumed error of $\pm 0.05$ mag
seems plausible and prudent. Col.~6 of Table~\ref{tab:ebl} shows the
perturbation to the best-fit value if all data-points are
systematically shifted by $\pm 0.05$ mag. Only for MIPS $24\mu$m does
this error dominate (Fig.~\ref{fig:errors}, green line).

\subsubsection{Spline fitting error}
The fitting process we adopt arbitrarily uses a 10-point
spline-fit. This was judged to be the lowest number of spline points
required to represent the data well. We repeat our analysis using an 8
or 12 point spline-fit and report the impact (Col.~7) on our best-fit
value using these alternative representations, i.e., $| \Delta
\rho_{L} | / 2$. In all cases except for $IRAC$-$4$, where other
issues have already been raised, the variation due to the fitting
process is negligible (Fig.~\ref{fig:errors}, red line).

\subsubsection{Poisson error \label{sec:errors}}
To assess the error arising from the uncertainty in the individual
data-points, we conduct 10,001 Monte-Carlo realizations where we
randomly perturb each data-point by its permissible error, assuming
the errors follow a Normal distribution.  For each sequence of
perturbations we re-fit the spline and extract the 16$^{th}$percentile
and the 84$^{th}$ percentile values for the luminosity density. The
uncertainty given in Col.~7 of Table~\ref{tab:ebl} is then $| \Delta
\rho_L |/2$.  We see that this error becomes dominant for data
longwards of MIPS $24\mu$m (Fig.~\ref{fig:errors}, blue line).

\subsubsection{Cosmic Variance error}
Finally we assess the error introduced by cosmic variance (or sample
variance), as discussed by \citet{dri10}. For each dataset shown in
Table.~\ref{tab:datasets}, we derive and assign a cosmic variance
estimate based on equation 4 in \citet{dri10}, using the appropriate
areas of the various datasets and some assumption of the likely
redshift range contributing to the counts (see
Table.~\ref{tab:datasets}). We conduct 10,001 Monte-Carlo
realizations, where we perturb each dataset by a random amount defined
by its cosmic variance (assuming the CV offset can be drawn from a
Normal distribution). We extract the 16$^{th}$percentile and the
84$^{th}$ percentiles, from which we derive an uncertainty, as, $|
\Delta \rho_L |/2$. We see from Fig.~\ref{fig:errors} (orange line)
  that this error dominates for most of the UV, optical, near-IR, and
  mid-IR bands.

\subsubsection{Combining errors}
Combining random and systematic errors is not entirely
straightforward, with some proponents advocating simply adding them
while others adding in quadrature, or keeping the systematic and
random errors separate. Here we take the most conservative approach of
adding the errors linearly, but provide the individual errors in
Table~\ref{tab:ebl} for those wishing to combine them in other
ways. The data points shown on subsequent figures are calculated
according to $\Delta_{\rm Total} = \Delta_{\rm Col.6} + \Delta_{\rm
  Col.7} + \Delta_{\rm Col.8} + \Delta_{\rm Col.9}$

Fig.~\ref{fig:errors} shows the contribution of each of the
  individual errors to our eIGL measurements and highlights the
  transition from being dominated by cosmic variance error at shorter
  wavelengths to random errors at longer wavelengths. At one
  wavelength ($IRAC$-$4$), we note that the fitting error itself
  dominates because of the complex shape of the data. One clear
  consequence from Fig.~\ref{fig:errors} is that deep, wide data are
  essential to reduce this error component, which should be readily
  achievable in the optical and near-IR, with the upcoming {\it
    Euclid} and the {\it Wide Field Infrared Space Telescope (WFIRST)}
  space-missions. Assuming calibration errors can be minimised, there
  is the distinct possibility of obtaining eIGL measurements to better
  than 1\% over the next decade.

\begin{figure*}

\plotone{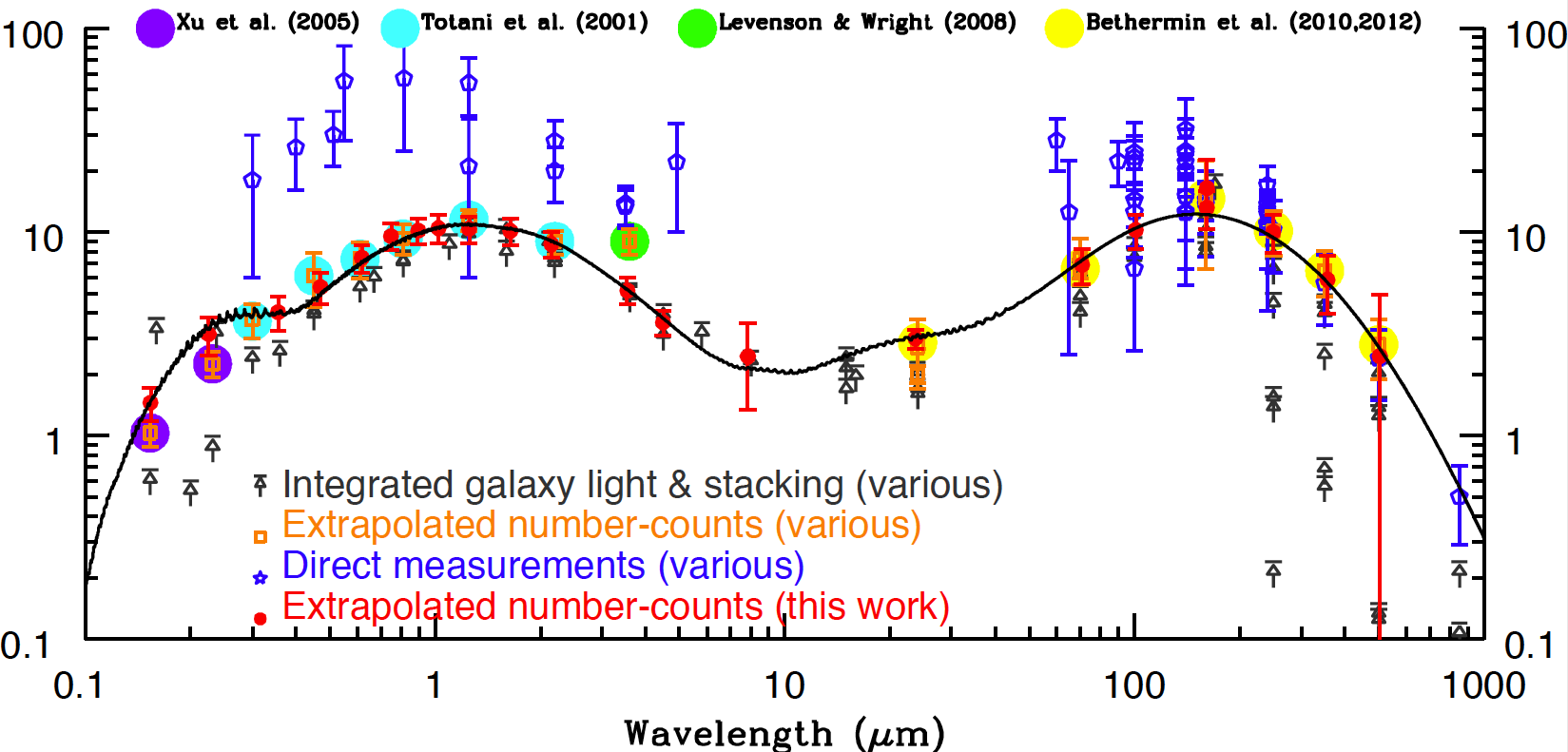}

\caption{Our measurement of the eIGL based on extrapolated
  number counts in each band compared to literature measurements taken
  from \citet{dwe13}. The black line depicts a phenomenological model
  from \citet{and16b}. The blue data points are attempted direct
  measurements which requires accurate modeling of both the Milky Way
  and Zodiacal light.  \label{fig:eblall}}
\end{figure*}

\begin{figure*}

\plotone{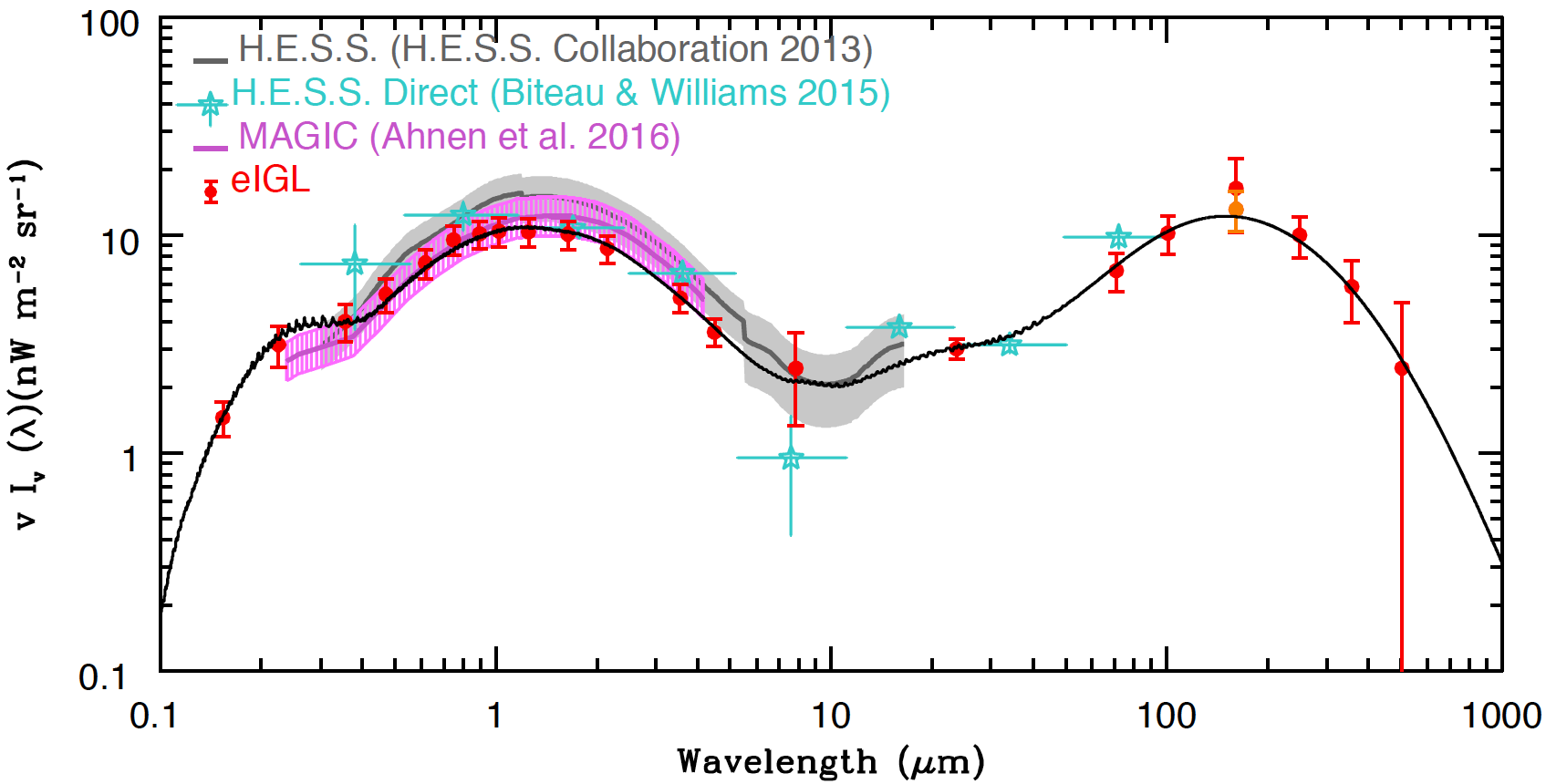}

\caption{Our measurements of the eIGL along with the model of Andrews
  et al. compared to the available VHE data from the H.E.S.S. and
  MAGIC Consortium (both of which use a pre-defined EBL model). The
  blue stars show the VHE constraint by \citet{bit15} which are
  independent of any pre-defined EBL model.
  \label{fig:vhe}}
\end{figure*}

\subsection{Comparison to previous EBL and eIGL measurements}
Fig.~\ref{fig:eblall} shows various IGL and EBL measurements as
reported over the past few years. Most of the data comes from the
comprehensive compilation by \citet{dwe13} (see their tables 3, 4 \& 5
for detailed references, although most appear here in the introduction
and text), plus more recent measurements by \citet{ash15}. We colour
code the data into three sets following Dwek \& Krennrich: lower
limits on the IGL (grey), IGL measurements based on various
extrapolated number counts (orange); and direct measurements of the
EBL via various methods (blue). Our new data is shown in red and
appears consistent with most previous eIGL measurements. Within the
eIGL data we only see one major discrepancy which is with the
\citet{lev08} value in the $IRAC$-$1$ band. We find a significantly
($\times 2$) lower value. \cite{lev08} also combine galaxy count data
from a number of sources and use two methods for deriving photometry,
profile fitting and apertures. Both require significant corrections
(upward shifts of $\sim40$\%). The value reported is from profile
fitting, however, we note that their corrected aperture value is
significantly lower ($5.9 \pm 1.0$ nW m$^{-2}$ sr$^{-1}$), and
consistent with our measurement ($5.2 \pm 0.8$ nW m$^{-2}$
sr$^{-1}$). We argue that, as our 3.6$\mu$m measurement lies in
between our $2.2\mu$m and $4.6\mu$m measurements, it is likely that
the \citet{lev08} profile-fit value is biased high.

In the far-UV and near-UV, we recover higher measurements than those
reported by \citet{xu05}, although formally at the $1\sigma$
error-limits.  However, this is nonetheless consistent with the data
shown in \citet{xu05} (see their figure~1) where their number counts
in both the far-UV and near-UV, to which their model is fitted, do
appear to fall systematically below the comparison datasets. See also
figure~6 of \citet{voy11} which shows an offset between the data of
\citet{xu05} and \citet{ham10}. Table.~2 of \citet{voy11} reports eIGL
measurements by \citet{voy11} and \cite{xu05} as well as \cite{gar00}
and earlier studies. Our value agrees well --- lies between --- the
two estimates provided by \citet{voy11}. In the near-UV our result is
dependent on an entirely different dataset, namely F225W observations
of the UVUDF. This base data is very different to the HST ACS SBC data
of \citet{voy11}, and agrees closely to the much earlier HST STIS data
of \citet{gar00}. We therefore conclude that the UV excess seen in our
data against the model is supported by three distinct datasets and
therefore likely real and significant.

In the far-IR, we see that our measurements mostly agree with those
previously reported. The one obvious outlier is the PACS 160$\mu$m
data, however we note (see Table~\ref{tab:ebl}) that a significant
amount of flux is coming from the extrapolation. In particular the
deepest data points from \citet{mag13} do include significant
completeness corrections. If we re-fit using only data with
completeness corrections at $<20$\% (their filled data points on their
figure~6), we recover a much more consistent value (see the second
entry in Table~\ref{tab:ebl} and orange data point on
Fig.~\ref{fig:eblall}). We therefore elect to adopt this revised data
point as the more robust estimate.

Most apparent from Fig.~\ref{fig:eblall} is the discrepancy in the
optical to near-IR between all the eIGL data (including our own), and
the direct measurements. This is in stark contrast to the far-IR where
the eIGL and EBL measurements agree within the specific errors. In the
case of the far-IR, the agreement is reassuring, and the much smaller
error bars on the eIGL measurements suggest that the eIGL route is the
more robust. Why then do we see such a discrepancy in the optical?
The model curve (black) shows the energy evolution model reported in
\citet{and16b}, which agrees closely with the eIGL data. Both the
model and the general consensus in the far-IR would therefore suggest
that the error may lie in the direct measurements, some of which do
concur with the eIGL estimates. It is worth noting that direct
measurements rely on a robust subtraction of the foreground light of
which there are two dominant sources: the Milky Way stellar population
and Zodiacal light (\citealp{hau01}; \citealp{mat06}). That the
discrepancy is most apparent at a wavelength comparable to the peak in
the solar spectrum also suggests that one or both of these foregrounds
is the source of the problem, and that either the Milky Way model or
the Zodiacal model have been underestimated.  One indication for the
latter over the former is the reanalysis of data taken between
1972---1974 at 0.44$\mu$m and 0.65$\mu$m by the {\it Pioneer~10/11}
spacecraft \citep{pioneer}. During that period the spacecraft were
approximately 4.66AU away from the Sun where the Zodiacal light
contribution should be negligible. Matsuoka found significantly lower
EBL measurements that are in agreement with our eIGL values.  We
advocate that given this information, it might be useful to adopt the
eIGL as the {\it de facto} measurements of the EBL, and use these to
help improve the Zodiacal light and inner Solar System dust model.

\subsection{Comparison to very high energy data}
Fig.~\ref{fig:vhe} shows the comparison of our eIGL data to three
  VHE datasets (as indicated).  The agreement is much better than with
  the direct estimates, and provides additional independent evidence
  that the direct estimates may be in error. Note that the
  H.E.S.S. and MAGIC datasets each adopt a pre-defined EBL model and
  solve for the normalisation, hence the slight shape discrepancy
  between the VHE and eIGL data is of no significance. Formally, the
  datasets overlap within the 1$\sigma$ errors, although the error
  range of the VHE data is fairly broad ($\times2$). As discussed in
  the introduction, the VHE data also comes with some caveats, in
  particular the assumption of the intrinsic shape of the Blazar
  spectra(um), the possibility of other interactions, e.g., with the
  intergalactic magnetic field or with PeV cascades. Nevertheless, the
  agreement is extremely encouraging and taken at face value suggests
  that our eIGL measurements are close to the underlying EBL values.

\subsection{Potential sources of missing light}
Before equating our eIGL measurement to the EBL we should
first acknowledge, in particular, the possible contributions from the
low surface brightness Universe: that from intra-cluster and
intra-group light \citep{zwi51}, also referred to as Intra-Halo Light
or IHL \citep{zem14}; and that from the epoch of reionisation
\citep{coo12}. 

\subsubsection{Low surface brightness galaxies}
The space-density of low surface brightness galaxies is currently
poorly constrained, however observations in the local group suggest
that the luminosity density is very much dominated by the Milky Way
and Andromeda. This picture is generally supported by our own estimates of
the low surface brightness population from the {\it HST} HDF \citep{dri99}
and the Millennium Galaxy Catalogue (see \citealp{lis03};
\citealp{dri05}).  Both studies explored the low surface brightness
Universe and, while finding numerous new systems, they ultimately contribute only
small amounts of additional light ($<20$\%; \citealp{dri99}). Studies
of rich clusters have also been very successful at finding
low surface brightness systems (e.g., \citealp{dav15};
\citealp{van15}), yet not in sufficient quantities to significantly
affect the total luminosity density, e.g., the thousand new galaxies
found in the Coma cluster by \citet{kod15} collectively add up to
just one extra L$^*$ galaxy.

Two final arguments can be made for a minimal amount of missing light
from low surface brightness galaxies based on the number-count and IGL
data itself.  Firstly, as each survey would have distinct surface
brightness cutoffs, any large population would be truncated at
different surface brightness levels leading to stark mismatches
between the distinct surveys. That the surveys overlap so well is a
strong argument for any missing population being relatively modest in
terms of their luminosity density. Secondly, any missing population of
galaxies will contain both optical emission from the starlight, and
dust emission from reprocessed starlight. Hence the consistency
between the far-IR EBL and eIGL can provide a constraint. To assess
this we compare our values of the eIGL to the direct EBL measurements
of \citet{fix98}, and derive (eIGL/EBL) ratios of 0.96, 0.97, 1.05 and
1.03 in 160, 250, 350 and 500$\mu$m bands for an average of 1.01,
i.e., on average 100\% of the direct EBL is ``resolved'' by our eIGL
measurements.

This therefore only gives no room for an upward adjustment for missing
``dusty'' galaxies. However we do acknowledge that the errors in both
our data and the Fixsen data are significant (typically 40\% per
band), and hence this close agreement must be somewhat
coincidental. Folding in the errors there is potentially room for an
$\sim19\%$ upward adjustment, i.e., $40\%/\sqrt{4}$, before
our eIGL exceeds the Fixsen EBL measurements by their reported
1$\sigma$ errors. Of course this argument assumes that the dust
properties of low surface brightness galaxies are consistent with
those of normal spiral galaxies, which may not be the case. A similar
conclusion, with regards missing galaxy light, was also reached by
\citet{tot01}, who specifically explored the potential impact of
surface brightness selection in deep Subaru number counts, and
concluded that any impact from missing low surface brightness
galaxies, via number-count modeling, was $<20$\% in the $BVIJL$ bands.
Our range of 20\% is comparable and hence we can adopt a possible
0---20\% upward adjustment range for missing low surface brightness
galaxies.

\subsubsection{Intra-cluster and intra-group light}
The case for the intra-cluster light is slightly less
clear-cut. \citet{mih05} shows spectacular images of nearby systems,
including Virgo, which typically contain between 10-20\% of the light
in a diffuse component. In the Frontiers' cluster A2744, \citet{mon14}
find that the ICL makes up only 6\% of the stellar mass. Similarly,
the study by \citet{pre14} also finds a relatively modest amount of
mass (8\%) in the CLASH-VLT cluster MACS J1206.2-0847. Earlier studies
of Coma, perhaps the most studied system, found significantly more
diffuse light, extending up to almost 50\% \citep{ber95}, and
simulations by \citet{rud11} suggest the ICL might contain anywhere
from 10-40\% of a rich cluster's total luminosity. Hence, studies of
the ICL could be used to argue for an upward adjustment of the
optical-only light of between 10-50\%. However it is important to
remember that rich clusters, such as Coma, the Frontier's, and CLASH
clusters are exceedingly rare \citep{eke05}, with less than 2\% of the
integrated galaxy light coming from $>10^{14.5}$M$_{\odot}$ haloes
(see \citealp{eke05} and \citealp{dri16b}).

In the absence of quality data, intuition can lead one in both
directions, as the fraction of diffuse light is likely to be a function
of the halo velocity dispersion, and the galaxy-galaxy interaction
velocity, duration, and frequency. Certainly evidence from the local
group suggests a fairly modest contribution, with the Magellanic
stream representing the only significant known source of diffuse
light. Furthermore the deep study of M96 (Leo I group) by
\citet{wat14}, failed to identify any significant intra-group light
to limits of 30.1 mag/sq arcsec. 

Intra-halo light will only affect the optical bands as it is dust free
due to the UV flux pervading the ICM destroying any dust particles. We
can therefore gauge the possible level by comparing our eIGL $g$ band
measurement to the EBL. As mentioned earlier, because of uncertainties
in the Zodiacal light model we cannot use most of the direct
estimates, however we can use the values provided by \cite{pioneer}
obtained from {\it Pioneer~10/11}. Using these we find eIGL/EBL ratios of
$0.68\pm0.23$ and $0.97\pm0.43$ in the $g$ and $r$ bands respectively
(with the errors dominated by the uncertainty in the {\it Pioneer}
estimates). The errors are large but suggest that the contribution
from the IHL (which can only be positive) lies in the range 3-32\% but
with a possible extreme upper limit of $\sim$35\% (i.e.,
$\frac{(0.68-0.23)+(0.97-0.43)}{2}/\sqrt{2}$) upward correction
of the optical and near-IR data (in-line with our earlier discussion
of the ICL).

\subsubsection{Reionisation}
Reionisation could potentially provide an additional diffuse photon
field in the near-IR range, i.e., where Ly$\alpha$ is redshifted
longwards of $1\mu$m.  Recent modeling by \citet{coo12} suggests the
flux of reionisation today might be in the range 0.1 - 0.3 nW m$^{-2}$
sr$^{-1}$ at $\sim 1.1\mu$m, i.e. a 3\% effect. This is well below our
quoted errors and so reionisation is unlikely to significantly impact
upon our measurements. However, it is worth noting that the
reionisation models are uncertain, and a more rigid upper-limit of
$\sim 2$ nW m$^{-2}$ sr$^{-1}$ at $1\mu$m was set by \cite{mad05},
based on arguments which related to the production of excessive metals
if the reionising flux was any stronger. This latter level is
plausibly detectable and could cause our eIGL to underestimate the EBL
by $\sim20$\% in the near-IR. Hence comparison of the near-IR eIGL to
direct EBL measurements could conceivably detect the reionisation
field.

\begin{figure*}

\plotone{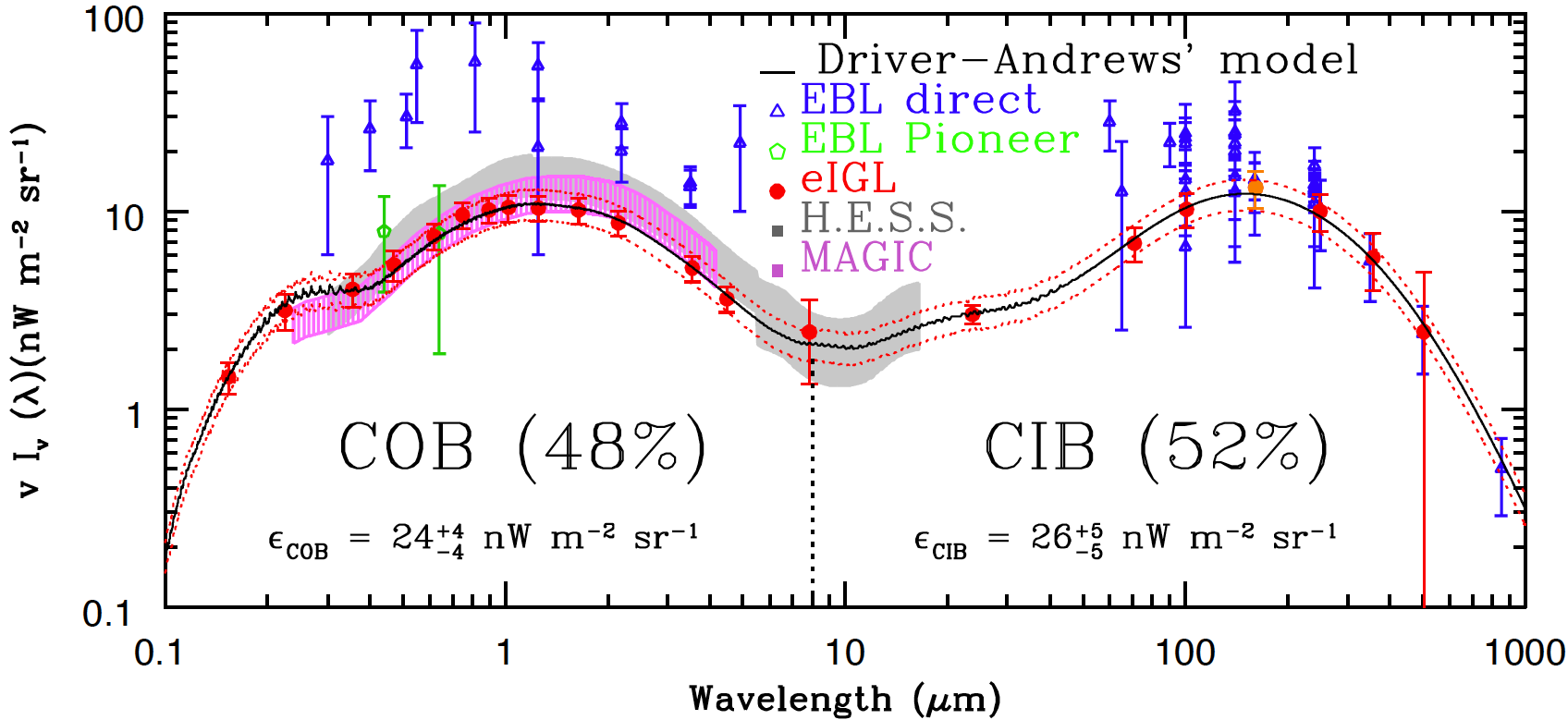}

\caption{Our derived eIGL data (red points), compared to terrestrial
  and near Solar System direct measurements (blue triangles), and the
  direct estimates from the {\it Pioneer} spacecraft at $>4.5AU$ (green
  circles). Also shown is the modified Driver-Andrews' model (black
  line) and the error bounds (red dashed lines) adopted for
  determining the uncertainty in the integrated COB and CIB
  measurements (as shown). \label{fig:eblplot2}}
\end{figure*}

\subsubsection{eIGL $\rightarrow$ EBL}
The eIGL and the EBL are, from the discussion above, slightly
different entities where the eIGL represents the sum of all radiation
from bound galaxies while the EBL includes, in addition, diffuse light
from the IHL and the epoch of reionisation.  However we have estimated
in comparison to the available direct EBL estimates that the eIGL
should match the EBL to within 0---35\% in UV-optical-near-IR
bandpasses and 0---20\% in far-IR bandpasses. These numbers are also
consistent with the discrepancy between our eIGL measurements and the
indirect VHE measurements. Specifically in $J$ band we find an eIGL
value of $10.38 \pm 1.52$ while the H.E.S.S. Collaboration find a
range for the EBL of 18.5----11.5 and the MAGIC team a range of
14.8---9.8. Formally our values are consistent. Again the means our
value is low by 45\% and 18\% but with no ($<1\sigma$ significance).
The take home message is more that the eIGL and VHE EBL data show
promising consistency, and if the errors in both can be reduced in
significance comparisons could potentially provide very interesting
constraints on the diffuse Universe.

\begin{figure*}

\epsscale{1.0}

\plotone{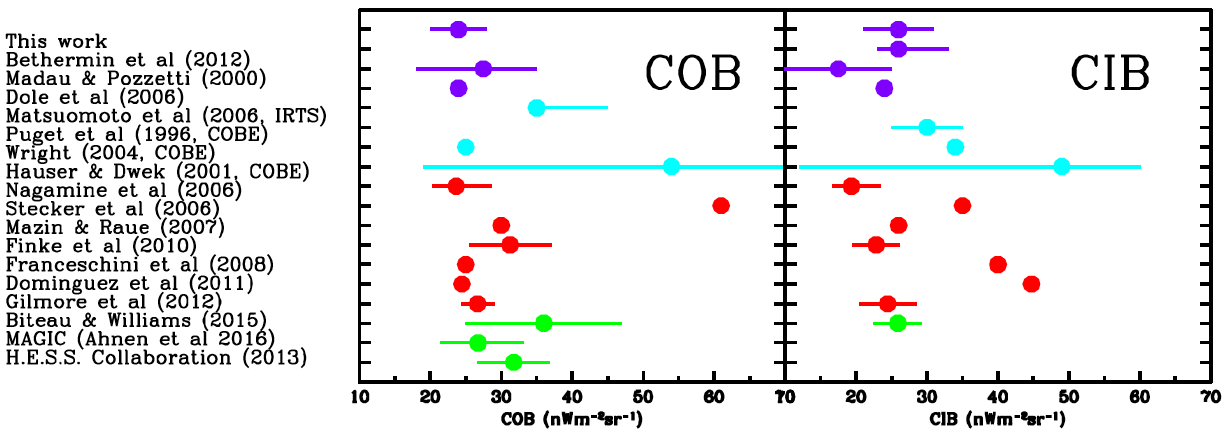}

\caption{A selection of available COB and CIB measurements which
  includes eIGL estimates (mauve), direct measurements (cyan),
  numerical models (red), and VHE results (green). Error ranges are
  shown if provided. \label{fig:compare}}
\end{figure*}

\subsection{The integrated energy  of the COB and CIB}
Finally, to determine the COB and CIB from our data we need to
identify an appropriate fitting function. In this case the most
straightforward option is to adopt a model which closely matches the
data. Shown on Figs.~\ref{fig:eblall} \&~\ref{fig:vhe} is a model
prediction from \cite{and16a} which is based on an update to the
two-component phenomenological model of \citet{dri13}. In this model,
we link spheroid formation to AGN activity, and adopted the axiom that
spheroid formation dominates at high-redshift. The variation of AGN
activity with redshift provides the shape, and the cosmic
star-formation history provides the normalisation, for the
star-formation history of spheroids only. The star-formation history
of discs is then the discrepancy between the total cosmic
star-formation history, and the spheroid star-formation history. With
the star-formation history of spheroids and discs defined, we use a
stellar population synthesis code and some underlying assumption of
the metallicity evolution (linear increase with star-formation), to
predict the cosmic spectral energy distribution at any epoch, and
compare to the available data at $z<0.1$ (see \citealp{dri13} for full
details). The model has now been developed to included obscured and
unobscured AGN, bolstering the UV flux, as well as dust reprocessing
and the model will be presented in detail in \citet{and16b}. 

In Fig.~\ref{fig:eblplot2} we again show our eIGL data compared to our
adopted model which provides a reasonable fit across the full
wavelength range shown. We perform a standard error-weight
$\chi^2$-minimisation of the model against the data to determine the
optimal normalisation and the $1\sigma$ error ranges on this
normalisation.  Note that we fit the EBL data to the COB and CIB
separately, and while the overall normalisations agree the recovered
$1\sigma$ errors are slightly broader for the CIB data (reflecting the
large associated errors). We now integrate the EBL model using the $R$
({\sc integrate}) function from $0.1$ to 8$\mu$m and $8$ to 1000$\mu$m
to obtain the total energy contained within the COB and CIB. We find
values of $24^{+4}_{-4}$ nW m$^{-2}$ sr$^{-1}$ and $26^{+5}_{-5}$ nW
m$^{-2}$ sr$^{-1}$ respectively, essentially a 48:52\% split. 

Fig.~\ref{fig:compare} shows some of the COB and CIB measurements
reported in the literature based on either integrated galaxy counts
(mauve), direct estimates (cyan), numerical models (red) or VHE data
(green). Our values agree well with previous eIGL estimates, and in
particular with the most recent CIB measurement of \citet{bet12}
($27^{+7}_{-3}$ nW m$^{-2}$ sr$^{-1}$). This should not be
particularly surprising as our CIB fits lean heavily on the B\'ethermin
source-count data, however the consistency in the measurement and
errors is reassuring. In general we do see a trend, that the eIGL
values are the lowest, the direct estimates the highest, and that the
VHE data is closer, but slightly higher, than the eIGL estimates. This
does imply that there may indeed be an additional diffuse component
(photon field) at the $\sim 20$\% level. As discussed above this could
plausibly be due to some combination of low surface brightness
galaxies, intra-halo light, and or any diffuse radiation from
reionisation. At the moment the errors are to broad to draw any firm
conclusion however, as observations improve in wide area imaging ({\it
  Euclid, WFIRST, LSST}), and in VHE capabilities, there is a strong
possibility of placing a meaningful constraint on this diffuse
component. In our analysis the dominant errors in the COB, at least,
are very much due to cosmic variance which are currently at the 5-10\%
level but can conceivably be reduced to below 1\% in the near future.
The normalised EBL model is available in machine readable format from
the ApJ online edition.

\section{Summary}
We have brought together a number of panchromatic datasets (GAMA,
COSMOS/G10, {\it HST} ERS, UVUDF and other mid and far-IR data) to
produce galaxy number counts which typically span over 10 magnitudes
and from the far-UV to the far-IR. Having homogenized the data, we
apply a consistent methodology to derive an integrated galaxy light
and an extrapolated galaxy-count light (eIGL) measurement. The method
avoids traditional galaxy number-count models and, as all datasets are
bounded in terms of the contribution to the IGL, are simply fit with a
10-point spline. Integrating the spline with or without extrapolation
then leads to a complete set of IGL measurements from the far-UV to
far-IR. Our error analysis includes four key components: a systematic
photometry and/or zero-point offset of $\pm 0.05$ mag in all datasets,
a re-fit based on 8 or 12 point splines, 10,001 Monte-Carlo
realizations of the random errors, and 10,001 Monte-Carlo realizations
of the cosmic variance estimates. We combine the errors linearly to
produce our final eIGL measurements, which are accurate to 2-30\%
depending on bandpass.

In comparison to previous data we generally agree with previous IGL
measurements, agree with direct measurements in the far-IR, but
disagree with direct measurements in the optical (see
Fig.~\ref{fig:eblall}, blue data points). We question whether the
Milky Way or Zodiacal light model requires revisiting for the direct
optical and near-IR measurements. In particular we note (see
Fig.~\ref{fig:eblplot2}) that the direct estimates from {\it Pioneer}
agree well with our eIGL estimates as do the constraints from very
high energy experiments, suggesting a possible issue with the inner
Solar System dust model.

We briefly acknowledge that the eIGL measurements could potentially
miss light from low surface brightness systems (0-20\%) and intra
cluster/group light (0-35\%). Insofar as studies exists evidence
suggests such emissions are likely small ($< 20$\%) and within our
quoted errors. However studies to further constrain both the
space-density of low surface brightness galaxies and the intra-halo
light would clearly be pertinent.

Finally we overlay the two-component model of \citet{dri13}, which now
includes AGN \citep{and16b} and find that we can explain the eIGL
distribution rather trivially in terms of a spheroid/AGN formation
phase ($z>1.5$), followed by disc formation ($z<1.5$). Using a
slightly modified version of the model as a fitting function we find
that the COB and CIB contain $24^{+4}_{-4}$ nW m$^{-2}$ sr$^{-1}$ and
$26^{+5}_{-5}$ nW m$^{-2}$ sr$^{-1}$ respectively, essentially a
48:52\% split.

Over the coming years with the advent of wide-field space-based
imaging, and in particular {\it Euclid} and {\it WFIRST}, we note the
  great potential to constraint the UV, optical and near-IR optical
  backgrounds to below 1\%.

\acknowledgements
We thank the anonymous referee for comments which led to improvements
in the paper, in particular the inclusion of cosmic variance errors.

We thank Adriano Fontana for providing the HUGS $K$ band number-count
data.  SPD acknowledges support from the Australian Research Council
via Discovery Project DP130103505.  RAW acknowledges support from NASA
JWST Interdisciplinary Scientist grant NAG5-12460 from GSFC.  LD
acknowledges support from the European Research Council in the form of
Advanced investigator Grant, COSMICISM and consolidated grant
CosmicDust.  SKA and AHW are supported by the Australian Governments'
Department of Industry Australian Postgraduate award.

GAMA is a joint European-Australasian project based around a
spectroscopic campaign using the Anglo-Australian Telescope. The GAMA
input catalogue is based on data taken from the Sloan Digital Sky
Survey and the UKIRT Infrared Deep Sky Survey. Complementary imaging
of the GAMA regions is being obtained by a number of independent
survey programmes including GALEX MIS, VST KiDS, VISTA VIKING, WISE,
Herschel-ATLAS, GMRT and ASKAP providing UV to radio coverage. GAMA is
funded by the STFC (UK), the ARC (Australia), the AAO, and the
participating institutions. The GAMA website is
\url{http://www.gama-survey.org/}. Based on observations made with ESO
Telescopes at the La Silla Paranal Observatory under programme ID
179.A-2004.

The COSMOS/G10 data is based on the spectroscopic catalogue of
\cite{dav15}, containing a reanalysis of the zCOSMOS \citep{lil07}
data obtained from observations made with ESO Telescopes at the La
Silla or Paranal Observatories under programme ID
175.A-0839. Photometric measurements are outlined in \citet{and16a} and
use data acquired as part of the Cosmic Evolution Survey (COSMOS)
project, derived using the \textsc{lambdar} software
\citep{wri16}. All data and derived products are available via the
COSMOS/G10 web portal: \url{http://ict.icrar.org/cutout/G10/}. This
web portal is hosted and maintained by funding from the International
Centre for Radio Astronomy Research (ICRAR) at the University of
Western Australia.

Data from a wide-range of facilities are included in this paper and we
wish to acknowledge the hard work and efforts of those involved in the
design, construction, operation and maintenance of these facilities
along with the science teams for having made their datasets readily
available: {\it GALEX, Hubble Space Telescope, Herschel Space
  Observatory, Spitzer, SDSS, CFHT, Subaru, Large Binocular Telescope,
  WISE, ESO (HAWK-I on UT4)}, the {\it Visible and Infrared Survey
  Telescope for Astronomy}, H.E.S.S. and MAGIC.

\appendix

\setcounter{table}{0}
\renewcommand{\thetable}{A\arabic{table}}

\section{Additional wavebands}
\begin{figure*}
\figurenum{A1}

\vspace{-1.0cm}

\epsscale{1.0}

\plotone{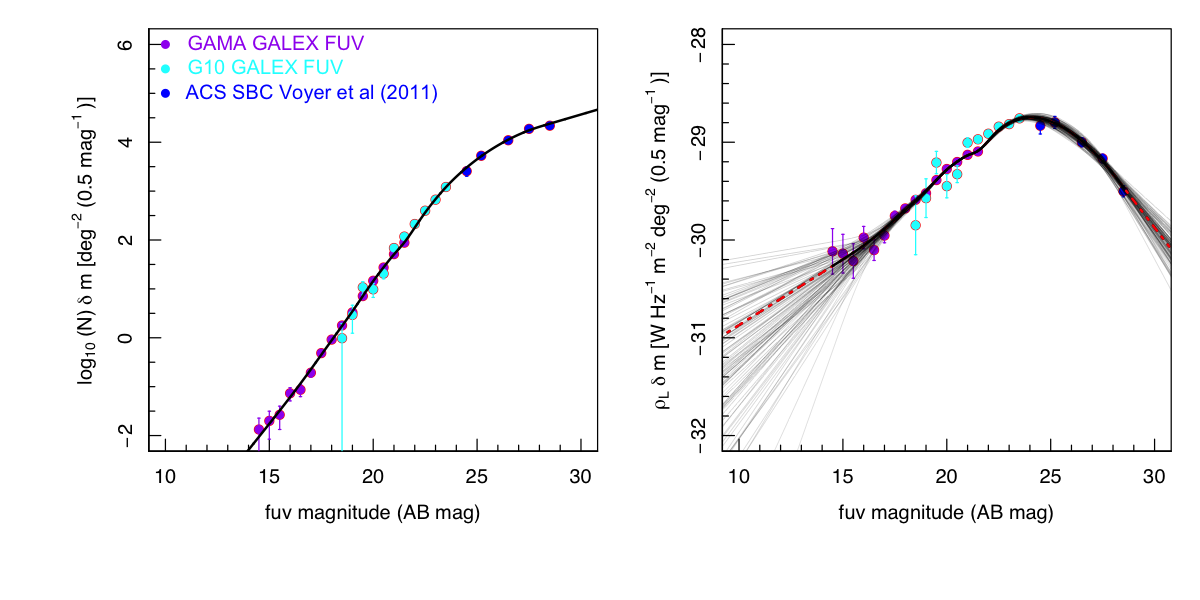}

\vspace{-1.0cm}

\plotone{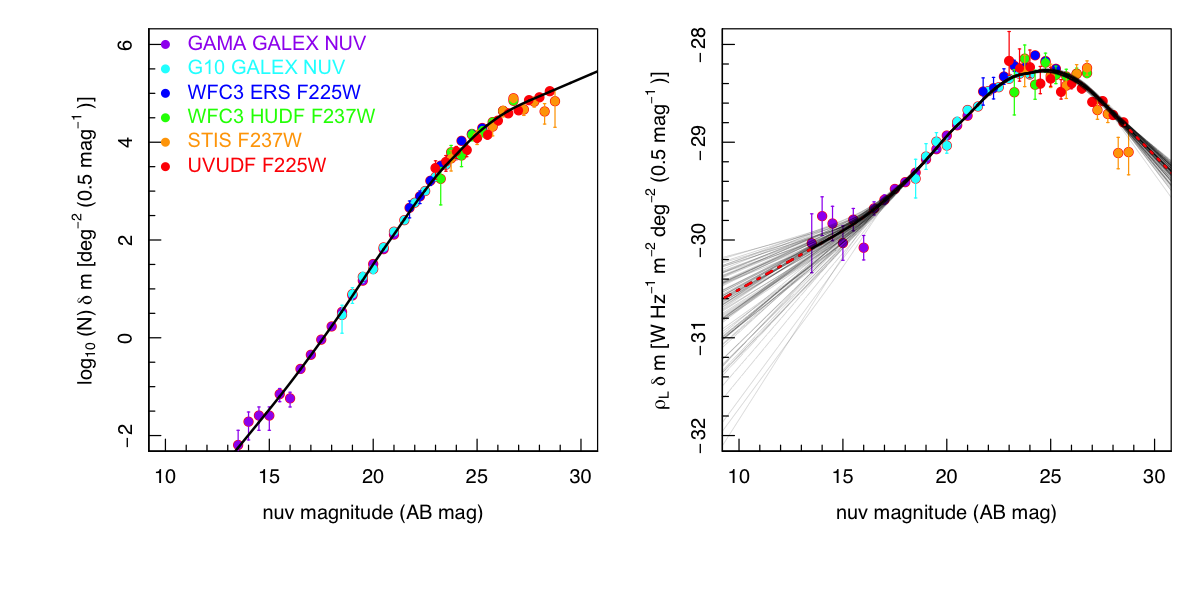}

\vspace{-1.0cm}

\plotone{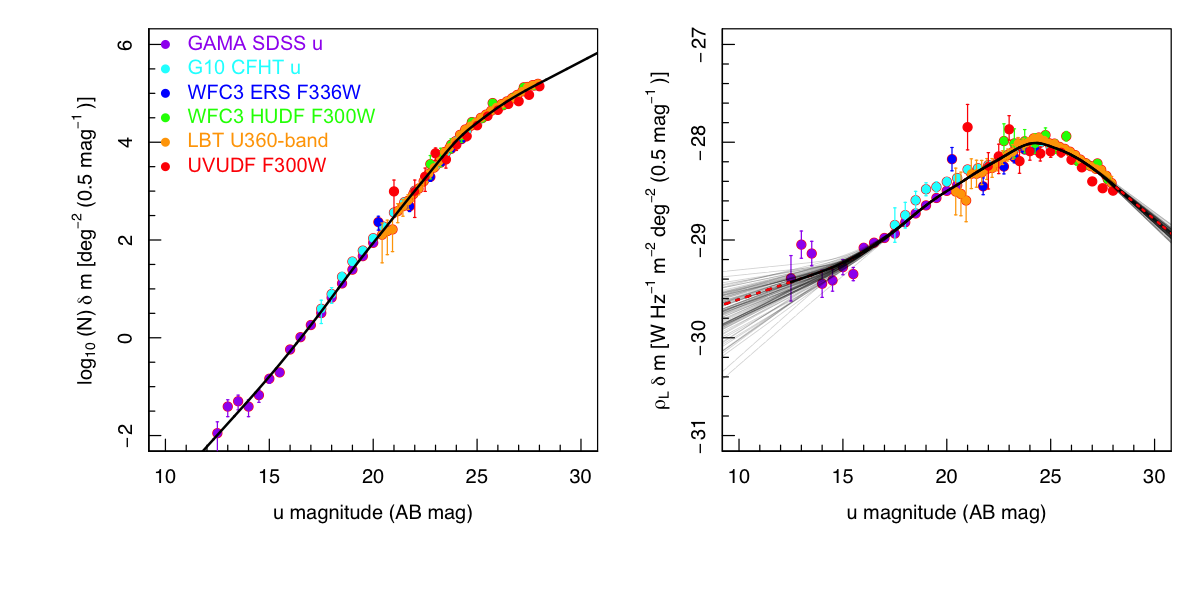}

\vspace{-1.0cm}

\caption{As for Fig.~\ref{fig:ebl} but for FUV, NUV and $u$ bands. \label{fig:ebl2}}

\end{figure*}

Figures~\ref{fig:ebl2} to \ref{fig:ebl3} shows the number count and
luminosity density figures for those datasets not shown in
Fig.~\ref{fig:ebl}, i.e., FUV/NUV, ugi, $ZYJHK$, $IRAC$-2\&4, MIPS24, PACS70/100/160
and SPIRE 350/500. The description of the lines, labels and key is as
for Fig.~\ref{fig:ebl}.

\begin{figure*}
\figurenum{A2}

\epsscale{1.0}

\vspace{-1.0cm}

\plotone{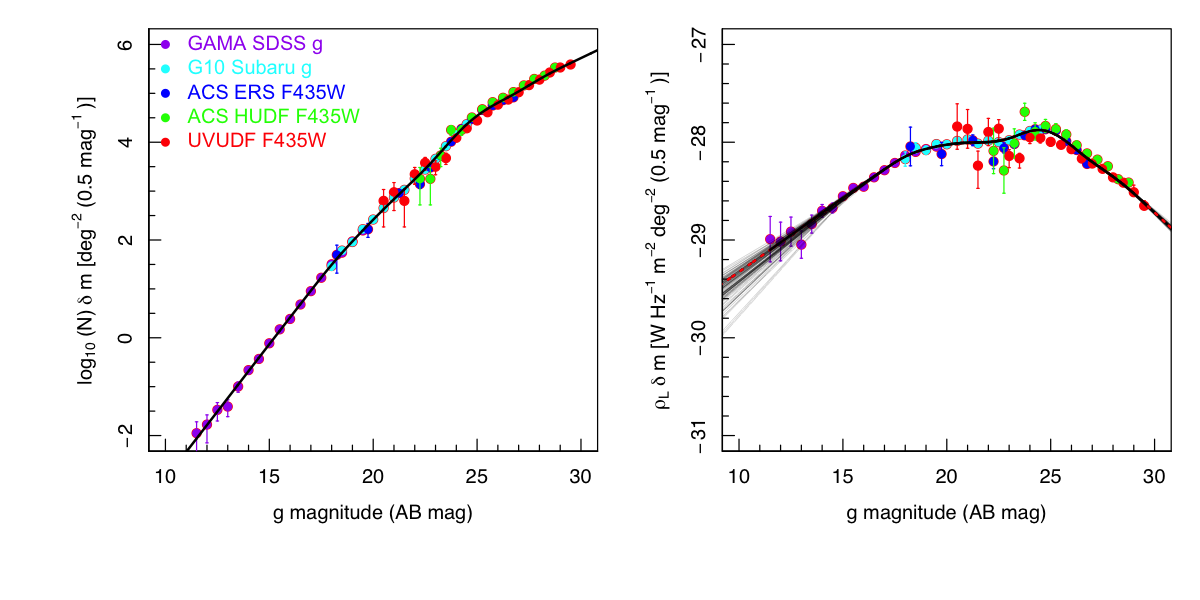}

\vspace{-1.0cm}

\plotone{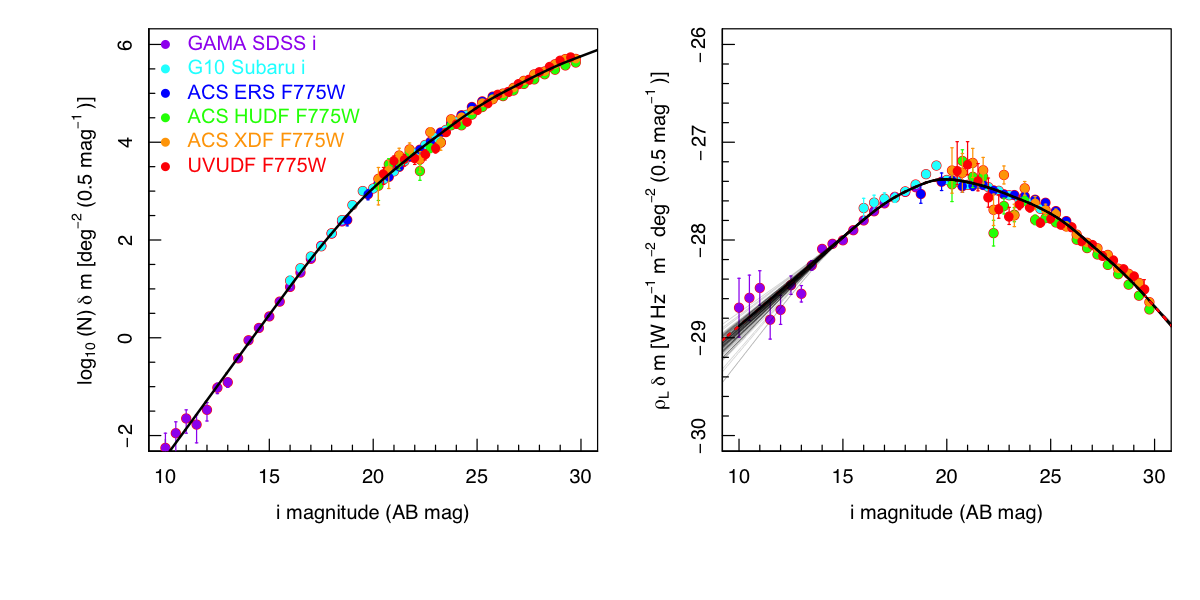}

\vspace{-1.0cm}

\plotone{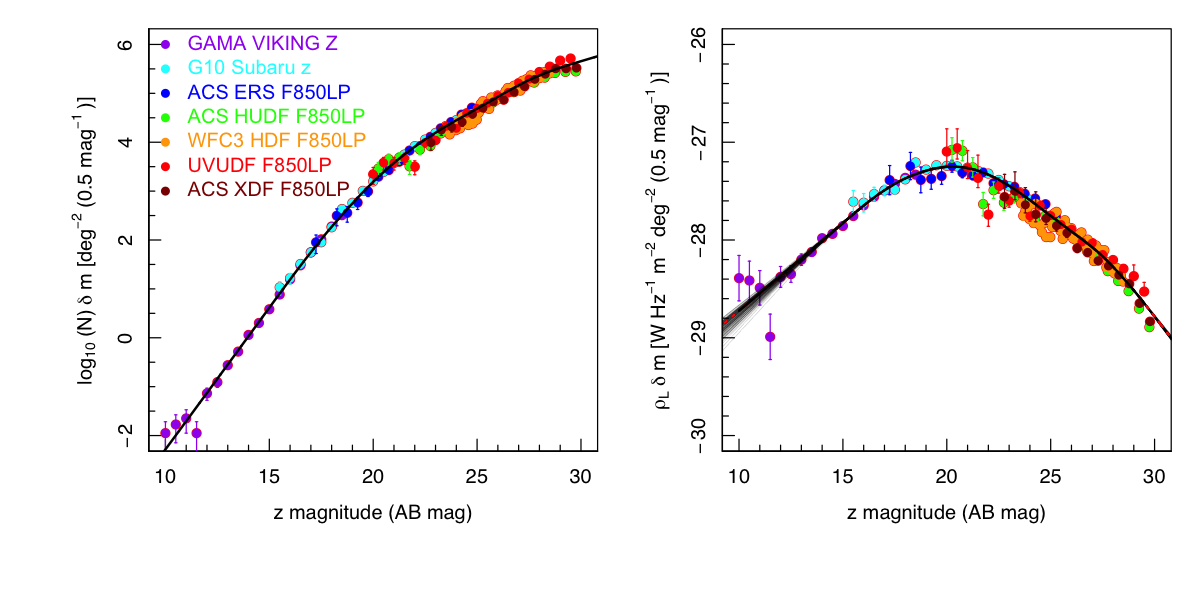}

\vspace{-1.0cm}

\caption{As for Fig.~\ref{fig:ebl} but for $g$, $i$ and $z$ bands.}

\end{figure*}

\begin{figure*}
\figurenum{A3}

\epsscale{1.0}

\vspace{-1.0cm}

\plotone{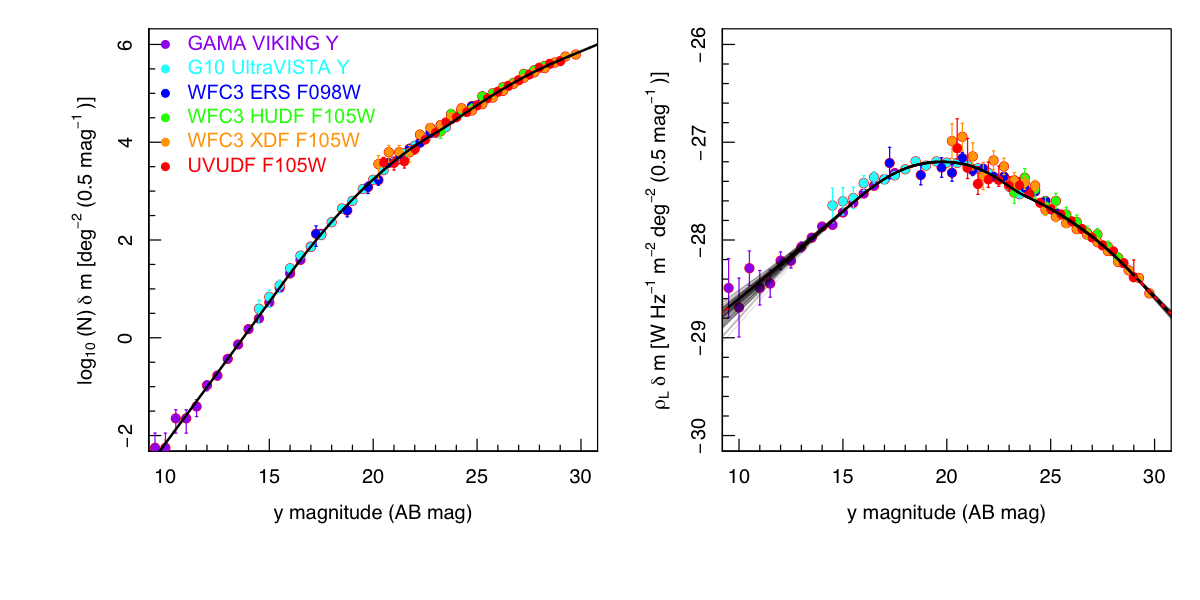}

\vspace{-1.0cm}

\plotone{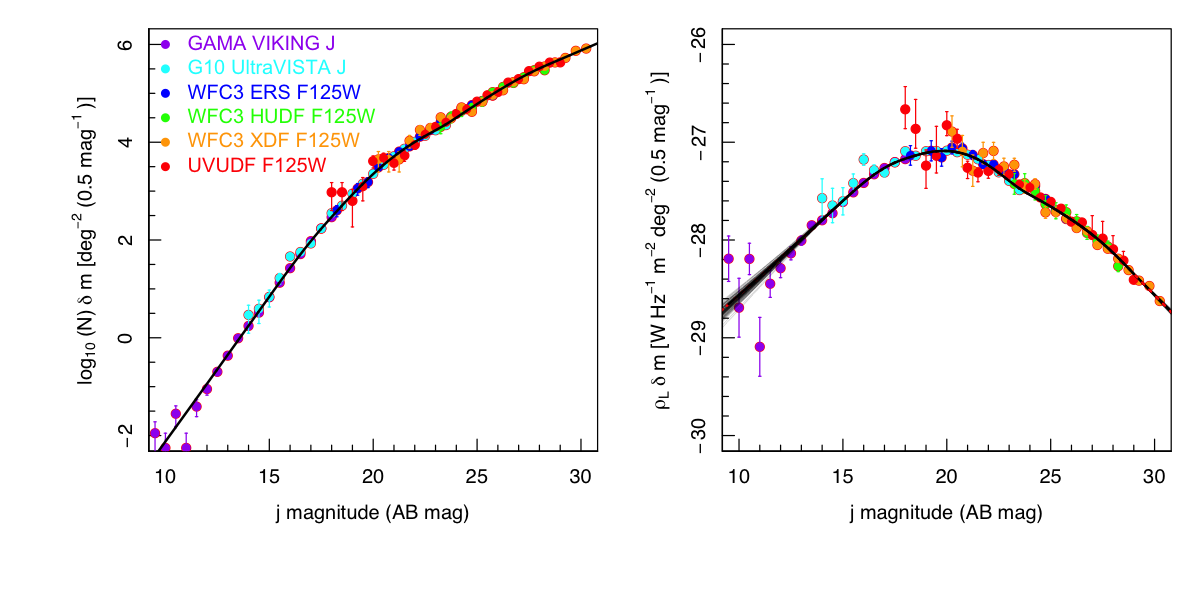}

\vspace{-1.0cm}

\plotone{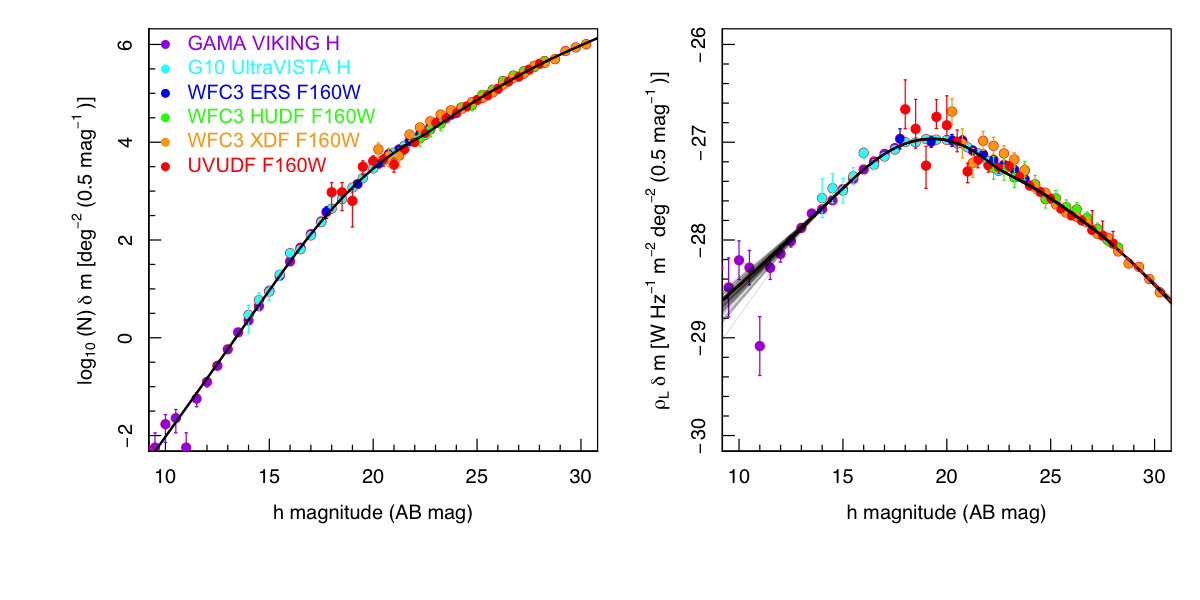}

\vspace{-1.0cm}

\caption{As for Fig.~\ref{fig:ebl} but for $Y$, $J$ and $H$ bands.}

\end{figure*}

\begin{figure*}
\figurenum{A4}

\epsscale{1.0}

\vspace{-1.0cm}

\plotone{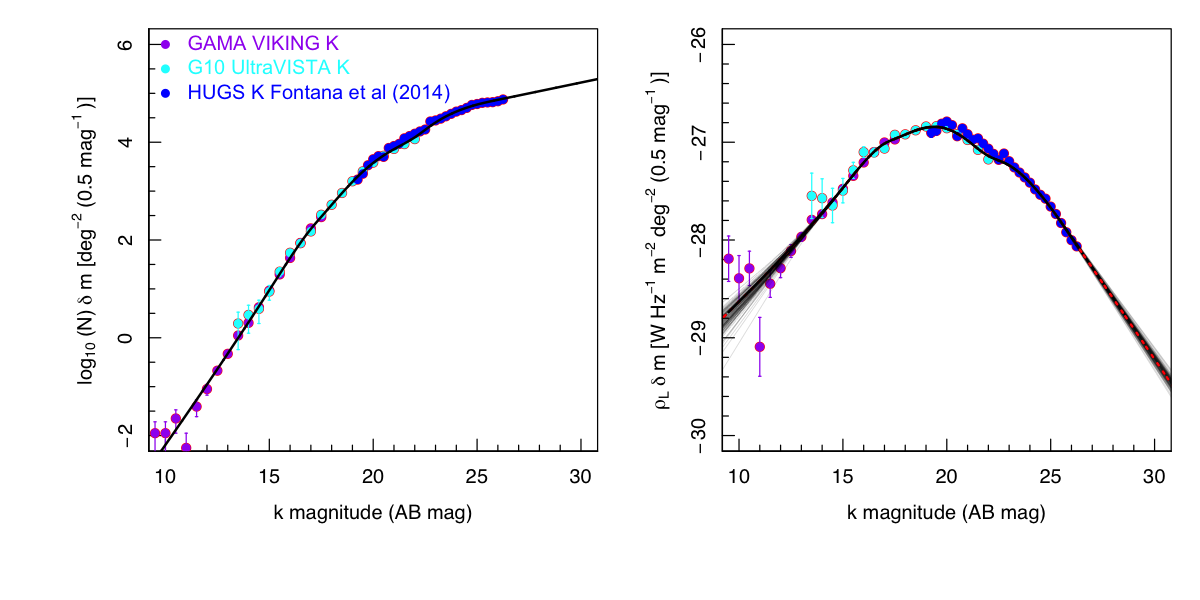}

\vspace{-1.0cm}

\plotone{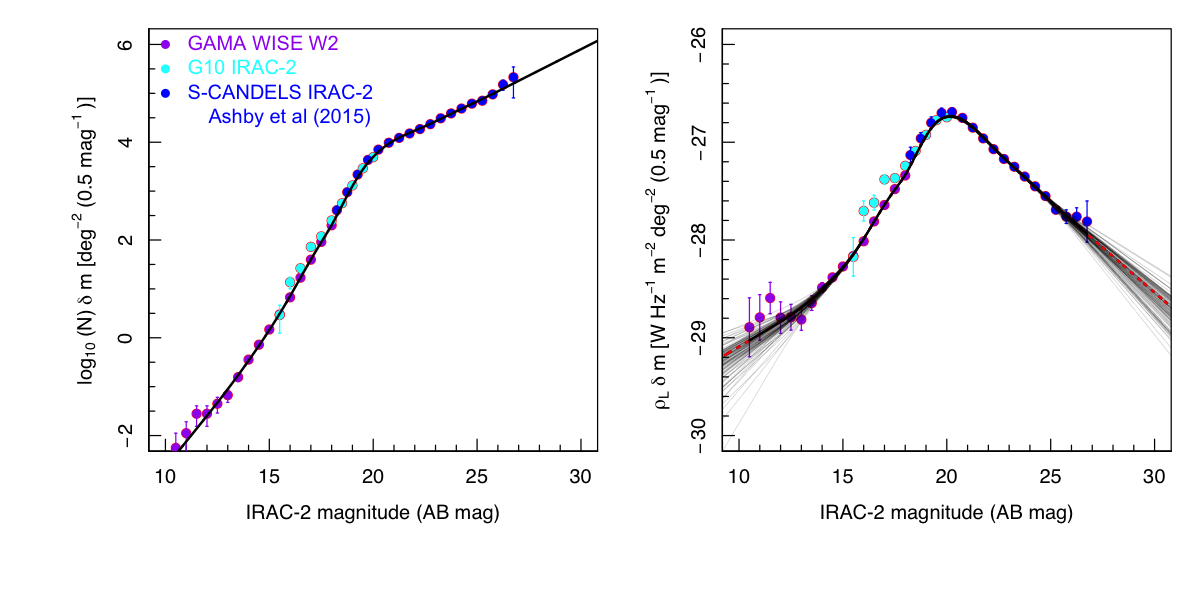}

\vspace{-1.0cm}

\plotone{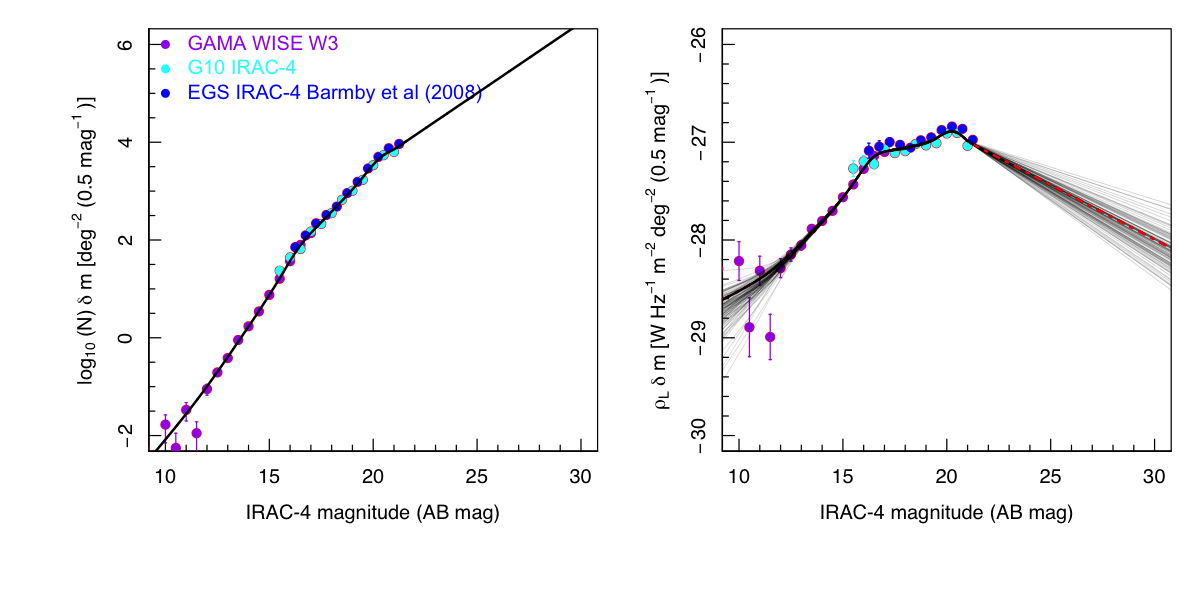}

\vspace{-1.0cm}

\caption{As for Fig.~\ref{fig:ebl} but for $K$, $IRAC$-$2$ and $IRAC$-$4$ bands. \label{fig:w3}}

\end{figure*}

\begin{figure*}
\figurenum{A5}

\epsscale{1.0}

\vspace{-1.0cm}

\plotone{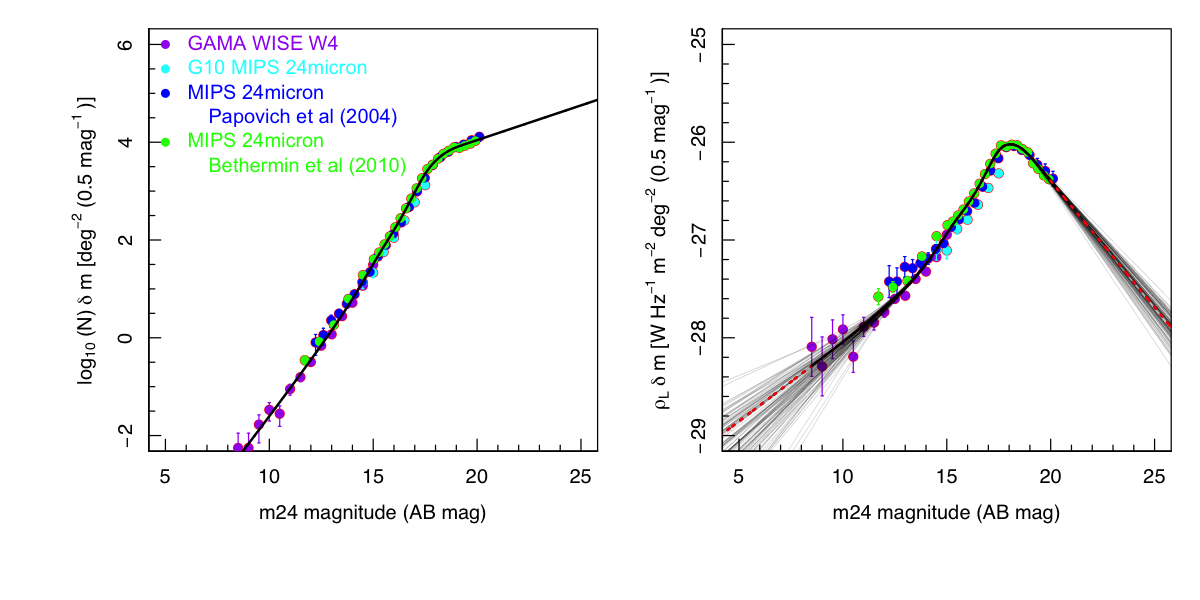}

\vspace{-1.0cm}

\plotone{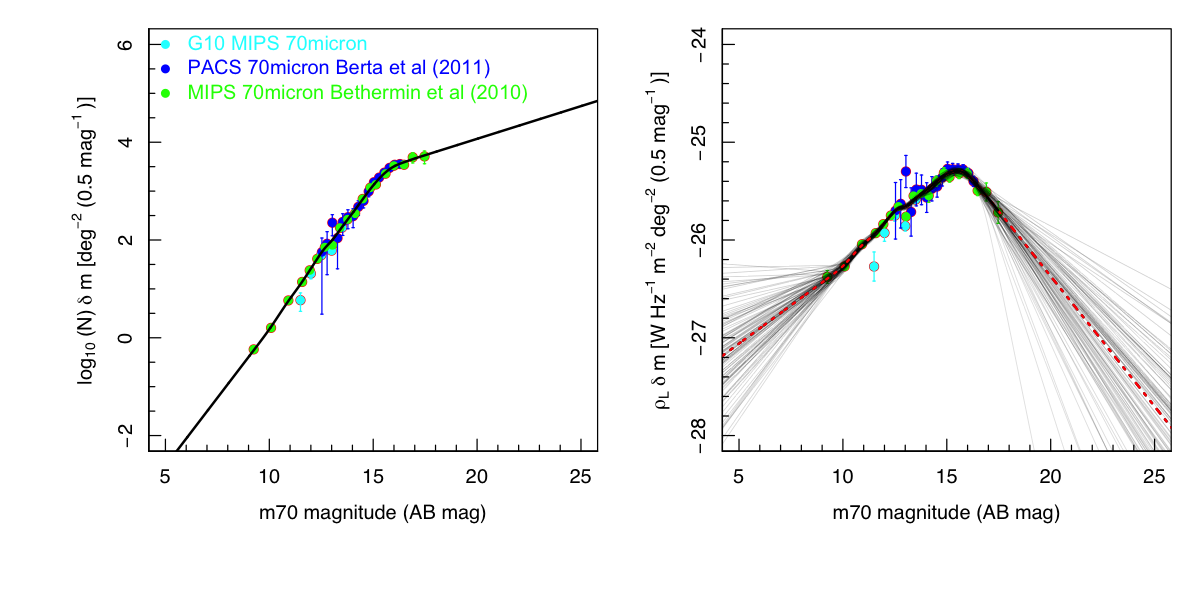}

\vspace{-1.0cm}

\plotone{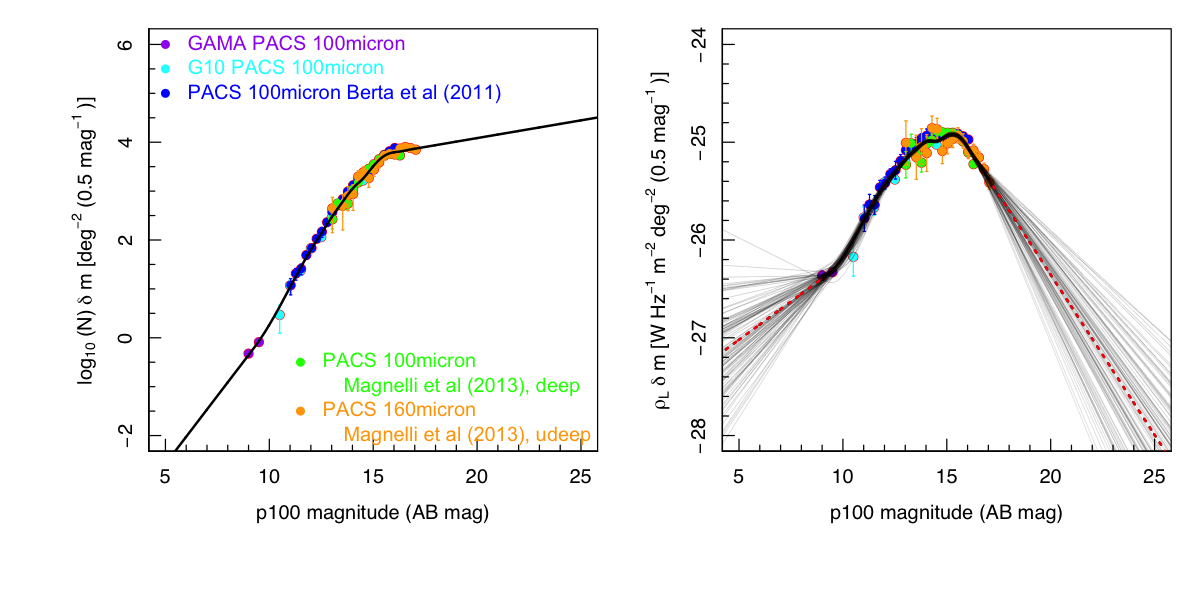}

\vspace{-1.0cm}

\caption{As for Fig.~\ref{fig:ebl} but for $24\mu$m, MIPS/PACS $70\mu $m and PACS $100\mu $m bands. \label{fig:pacs}}

\end{figure*}

\begin{figure*}
\figurenum{A6}

\epsscale{1.0}

\vspace{-1.0cm}

\plotone{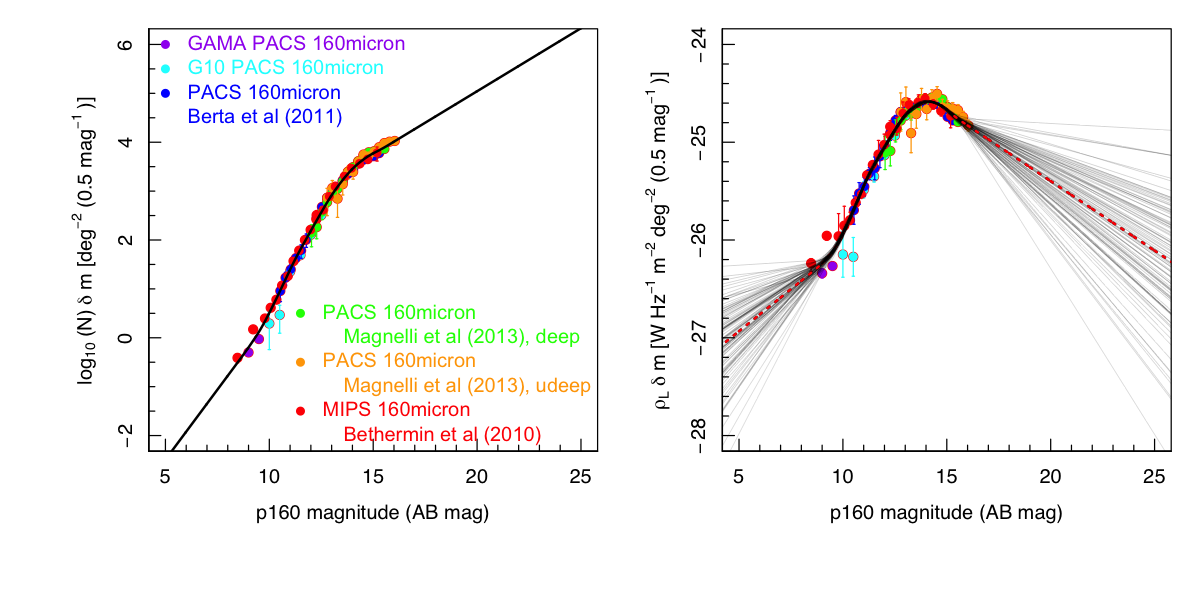}

\vspace{-1.0cm}

\plotone{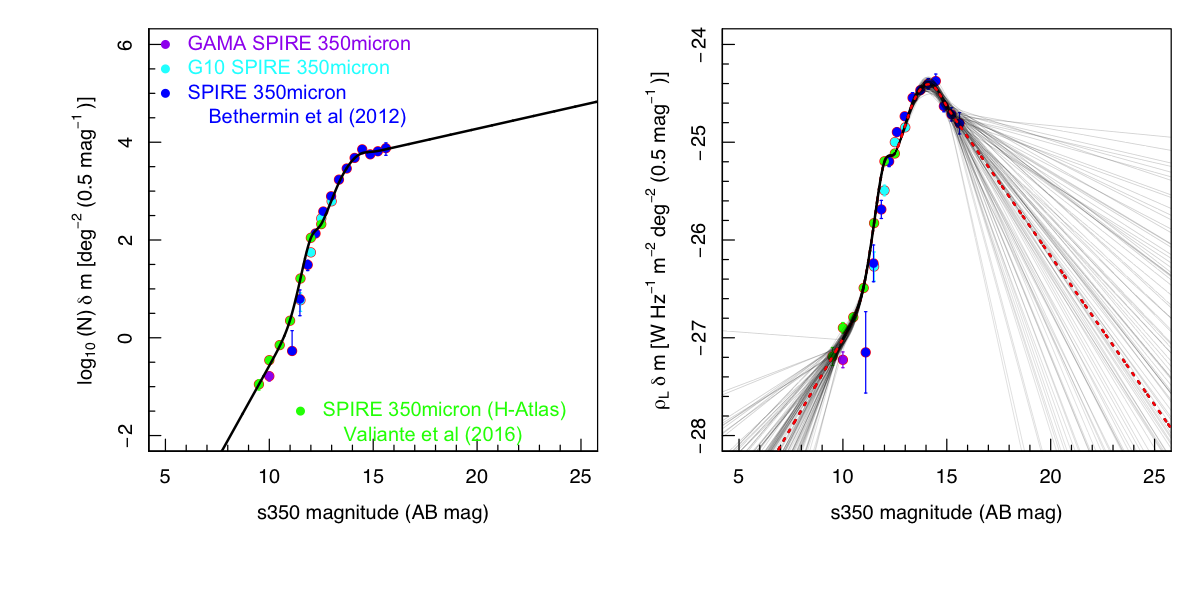}

\vspace{-1.0cm}

\plotone{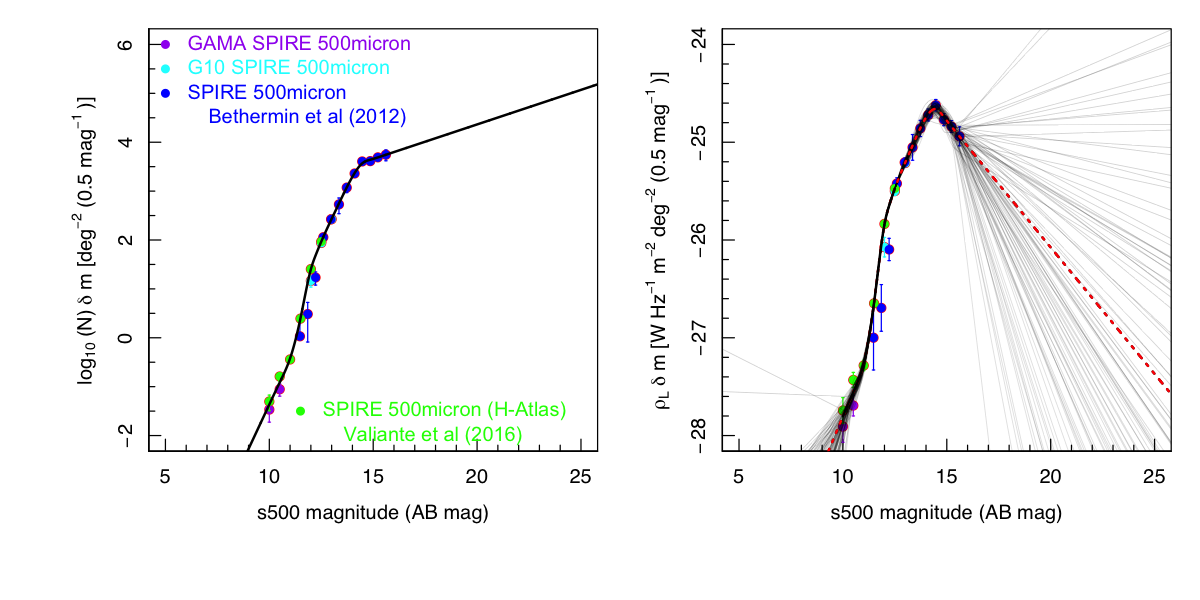}

\vspace{-1.0cm}

\caption{As for Fig.~\ref{fig:ebl} but for PACS $160 \mu $m, SPIRE $350\mu $m and SPIRE $500\mu $m bands. \label{fig:ebl3}}

\end{figure*}

\clearpage




\begin{thebibliography}{}

\bibitem[Aharonian et al.(2006)]{aha06} Aharonian, F., Akhperjanian, A.~G., Bazer-Bachi, A.~R., et al.\ 2006, \nat, 440, 1018 

\bibitem[Ahnen et al.(2016)]{ahn16} Ahnen, M.~L., Ansoldi, S., Antonelli, L.~A., et al.\ 2016, arXiv:1602.05239 

\bibitem[Alexander et al.(2005)]{ale05} Alexander, D.~M., 
Bauer, F.~E., Chapman, S.~C., et al.\ 2005, \apj, 632, 736 

\bibitem[Andrews et al.(2016a)]{and16a} Andrews, S.~K., et al.\ 2016a, \mnras, submitted

\bibitem[Andrews et al.(2016b)]{and16b} Andrews, S.~K., et al.\ 2016b, \mnras, in prep.

\bibitem[Ashby et al.(2015)]{ash15} Ashby, M.~L.~N., Willner, 
S.~P., Fazio, G.~G., et al.\ 2015, \apjs, 218, 33 

\bibitem[Barmby et al.(2008)]{bar08} Barmby, P., Huang, 
J.-S., Ashby, M.~L.~N., et al.\ 2008, \apjs, 177, 431 

\bibitem[Bernstein et al.(1995)]{ber95} Bernstein, G.~M., 
Nichol, R.~C., Tyson, J.~A., Ulmer, M.~P., 
\& Wittman, D.\ 1995, \aj, 110, 1507 

\bibitem[Bernstein(2007)]{ber07} Bernstein, R.~A.\ 2007, 
\apj, 666, 663 

\bibitem[Bernstein et al.(2002)]{ber02} Bernstein, R.~A., 
Freedman, W.~L., \& Madore, B.~F.\ 2002, \apj, 571, 56 

\bibitem[Berta et 
al.(2011)]{ber11} Berta, S., Magnelli, B., Nordon, R., et al.\ 2011, \aap, 532, A49 

\bibitem[Bertin \& Arnouts (1996)]{ber96} Bertin, E., Arnouts, S., 1996, A\&AS, 117, 393

\bibitem[B{\'e}thermin et 
al.(2010)]{bet10} B{\'e}thermin, M., Dole, H., Beelen, A., \& Aussel, H.\ 2010, \aap, 512, A78 

\bibitem[B\'ethermin et 
al.(2012)]{bet12} B{\'e}thermin, M., Le Floc'h, E., Ilbert, O., et al.\ 2012, \aap, 542, A58 

\bibitem[Biteau \& Williams(2015)]{bit15} Biteau, J., \& Williams, D.~A.\ 2015, \apj, 812, 60 

\bibitem[Cambr{\'e}sy et al.(2001)]{cam01} Cambr{\'e}sy, L., 
Reach, W.~T., Beichman, C.~A., \& Jarrett, T.~H.\ 2001, \apj, 555, 563 

\bibitem[Capak et al.(2007)]{cap07} Capak, P., Aussel, H., 
Ajiki, M., et al.\ 2007, \apjs, 172, 99 

\bibitem[Chabrier(2003)]{cha03} Chabrier, G.\ 2003, \pasp, 
115, 763 

\bibitem[Cluver et al.(2014)]{clu14} Cluver, M.~E., Jarrett, 
T.~H., Hopkins, A.~M., et al.\ 2014, \apj, 782, 90 

\bibitem[Cooray et al.(2012)]{coo12} Cooray, A., Gong, Y., 
Smidt, J., \& Santos, M.~G.\ 2012, \apj, 756, 92 


\bibitem[Dale et al.(2014)]{dal14} Dale, D.~A., Helou, G., 
Magdis, G.~E., et al.\ 2014, \apj, 784, 83 

\bibitem[de Oliveira-Costa et al.(2008)]{deo08} de 
Oliveira-Costa, A., Tegmark, M., Gaensler, B.~M., et al.\ 2008, \mnras, 
388, 247 

\bibitem[Dole et 
al.(2006)]{dol06} Dole, H., Lagache, G., Puget, J.-L., et al.\ 2006, \aap, 451, 417 

\bibitem[Davies et al.(2015)]{dav15} Davies, L.~J.~M., 
Driver, S.~P., Robotham, A.~S.~G., et al.\ 2015, \mnras, 447, 1014 

\bibitem[Davies et al.(2016)]{dav16} Davies, J.~I., Davies, 
L.~J.~M., \& Keenan, O.~C.\ 2016, \mnras, 456, 1607 

\bibitem[Dom{\'{\i}}nguez et al.(2011)]{dom11} 
Dom{\'{\i}}nguez, A., Primack, J.~R., Rosario, D.~J., et al.\ 2011, \mnras, 
410, 2556 

\bibitem[Driver et al.(2016a)]{dri16a} Driver, S.~P., Wright, 
A.~H., Andrews, S.~K., et al.\ 2016, \mnras, 455, 3911 

\bibitem[Driver et al.(2013)]{dri13} Driver, S.~P., Robotham, 
A.~S.~G., Bland-Hawthorn, J., et al.\ 2013, \mnras, 430, 2622 

\bibitem[Driver et al.(2012)]{dri12} Driver, S.~P., Robotham, 
A.~S.~G., Kelvin, L., et al.\ 2012, \mnras, 427, 3244 

\bibitem[Driver et al.(2011)]{dri11} Driver, S.~P., Hill, 
D.~T., Kelvin, L.~S., et al.\ 2011, \mnras, 413, 971 

\bibitem[Driver 
\& Robotham(2010)]{dri10} Driver, S.~P., \& Robotham, A.~S.~G.\ 2010, \mnras, 407, 2131 
 
\bibitem[Driver et al.(2008)]{dri08} Driver, S.~P., Popescu, 
C.~C., Tuffs, R.~J., et al.\ 2008, \apjl, 678, L101 

\bibitem[Driver et al.(2007)]{dri07} Driver, S.~P., Popescu, 
C.~C., Tuffs, R.~J., et al.\ 2007, \mnras, 379, 1022 

\bibitem[Driver(1999)]{dri99} Driver, S.~P.\ 1999, \apjl, 
526, L69 

\bibitem[Driver et al.(2005)]{dri05} Driver, S.~P., Liske, 
J., Cross, N.~J.~G., De Propris, R., \& Allen, P.~D.\ 2005, \mnras, 360, 81 
 
\bibitem[Driver et al.(2016b)]{dri16b} Driver, S.P., et al.\ 2016, \mnras, in prep.

\bibitem[Dunne et al.(2011)]{dun11} Dunne, L., Gomez, H.~L., 
da Cunha, E., et al.\ 2011, \mnras, 417, 1510 

\bibitem[Dwek 
\& Krennrich(2013)]{dwe13} Dwek, E., \& Krennrich, F.\ 2013, Astroparticle Physics, 43, 112 

\bibitem[Dwek 
\& Arendt(1998)]{dwe98a} Dwek, E., \& Arendt, R.~G.\ 1998, \apjl, 508, L9 

\bibitem[Dwek et al.(1998)]{dwe98b} Dwek, E., Arendt, R.~G., 
Hauser, M.~G., et al.\ 1998, \apj, 508, 106 

\bibitem[Eales et al.(2010)]{eal10} Eales, S., Dunne, L., 
Clements, D., et al.\ 2010, \pasp, 122, 499 

\bibitem[Eke et al.(2005)]{eke05} Eke, V.~R., Baugh, C.~M., 
Cole, S., et al.\ 2005, \mnras, 362, 1233 

\bibitem[Finke et al.(2010)]{fin10} Finke, J.~D., Razzaque, 
S., \& Dermer, C.~D.\ 2010, \apj, 712, 238 

\bibitem[Fixsen et al.(1998)]{fix98} Fixsen, D.~J., Dwek, E., 
Mather, J.~C., Bennett, C.~L., \& Shafer, R.~A.\ 1998, \apj, 508, 123 

\bibitem[Fontana et 
al.(2014)]{fon14} Fontana, A., Dunlop, J.~S., Paris, D., et al.\ 2014, \aap, 570, A11 

\bibitem[Franceschini et al.(2008)]{fra08} Franceschini, A., Rodighiero, G., \& Vaccari, M.\ 2008, \aap, 487, 837 

\bibitem[Frayer et al.(2009)]{fra09} Frayer, D.~T., Sanders, 
D.~B., Surace, J.~A., et al.\ 2009, \aj, 138, 1261 

\bibitem[Gardner et al.(2000)]{gar00} Gardner, J.~P., Brown, 
T.~M., \& Ferguson, H.~C.\ 2000, \apjl, 542, L79 

\bibitem[Gilmore et al.(2012)]{gil12} Gilmore, R.~C., Somerville,
  R.~S., Primack, J.~R., \& Dom{\'{\i}}nguez, A.\ 2012, \mnras, 422,
  3189

\bibitem[Grazian et 
al.(2009)]{gra09} Grazian, A., Menci, N., Giallongo, E., et al.\ 2009, \aap, 505, 1041 

\bibitem[H.E.S.S.~Collaboration et al.(2013)]{abr13}
  H.E.S.S.~Collaboration, Abramowski, A., Acero, F., et al.\ 2013,
  \aap, 550, A4

\bibitem[Hammer et al.(2010)]{ham10} Hammer, D., Verdoes 
Kleijn, G., Hoyos, C., et al.\ 2010, \apjs, 191, 143 

\bibitem[Hauser et al.(1998)]{hau98} Hauser, M.~G., Arendt, 
R.~G., Kelsall, T., et al.\ 1998, \apj, 508, 25 

\bibitem[Hauser 
\& Dwek(2001)]{hau01} Hauser, M.~G., \& Dwek, E.\ 2001, \araa, 39, 249 

\bibitem[Hill et al.(2011)]{hil11} Hill, D.~T., Kelvin, 
L.~S., Driver, S.~P., et al.\ 2011, \mnras, 412, 765 

\bibitem[Hopkins et al.(2013)]{hop13} Hopkins, A.~M., Driver, 
S.~P., Brough, S., et al.\ 2013, \mnras, 430, 2047 

\bibitem[Hopwood et al.(2010)]{hop10} Hopwood, R., Serjeant, 
S., Negrello, M., et al.\ 2010, \apjl, 716, L45 

\bibitem[Inoue et al.(2013)]{ino13} Inoue, Y., Inoue, S., 
Kobayashi, M.~A.~R., et al.\ 2013, \apj, 768, 197 

\bibitem[Jarrett et al.(2011)]{jar11} Jarrett, T.~H., Cohen, 
M., Masci, F., et al.\ 2011, \apj, 735, 112 

\bibitem[Jauzac et 
al.(2011)]{jau11} Jauzac, M., Dole, H., Le Floc'h, E., et al.\ 2011, \aap, 525, A52 

\bibitem[Kashlinsky(2006)]{kas06} Kashlinsky, A.\ 2006, \nar, 
50, 208 

\bibitem[Keenan et al.(2010)]{kee10} Keenan, R.~C., Barger, 
A.~J., Cowie, L.~L., \& Wang, W.-H.\ 2010, \apj, 723, 40 

\bibitem[Khaire 
\& Srianand(2015)]{kha15} Khaire, V., \& Srianand, R.\ 2015, \apj, 805, 33 

\bibitem[Koda et al.(2015)]{kod15} Koda, J., Yagi, M., 
Yamanoi, H., \& Komiyama, Y.\ 2015, \apjl, 807, L2 

\bibitem[Krick et al.(2009)]{kri09} Krick, J.~E., Surace, 
J.~A., Thompson, D., et al.\ 2009, \apjs, 185, 85 

\bibitem[Lagache et 
al.(2005)]{lag05} Lagache, G., Puget, J.-L., \& Dole, H.\ 2005, \araa, 43, 727 

\bibitem[Lagache et 
al.(1999)]{lag99} Lagache, G., Abergel, A., Boulanger, F., D{\'e}sert, F.~X., \& Puget, J.-L.\ 1999, \aap, 344, 322 

\bibitem[Levenson 
\& Wright(2008)]{lev08} Levenson, L.~R., \& Wright, E.~L.\ 2008, \apj, 683, 585 

\bibitem[Levenson et al.(2010)]{lev10} Levenson, L., Marsden, 
G., Zemcov, M., et al.\ 2010, \mnras, 409, 83 

\bibitem[Lilly et al.(2007)]{lil07} Lilly, S.~J., Le
  F{\`e}vre, O., Renzini, A., et al.\ 2007, \apjs, 172, 70

\bibitem[Liske et al.(2003)]{lis03} Liske, J., Lemon, D.~J., 
Driver, S.~P., Cross, N.~J.~G., \& Couch, W.~J.\ 2003, \mnras, 344, 307 

\bibitem[Liske et al.(2015)]{lis15} Liske, J., Baldry, I.~K., 
Driver, S.~P., et al.\ 2015, \mnras, 452, 2087 

\bibitem[Lutz et 
al.(2011)]{lut11} Lutz, D., Poglitsch, A., Altieri, B., et al.\ 2011, \aap, 532, A90 

\bibitem[Madau 
\& Pozzetti(2000)]{mad00} Madau, P., \& Pozzetti, L.\ 2000, \mnras, 312, L9 

\bibitem[Madau 
\& Silk(2005)]{mad05} Madau, P., \& Silk, J.\ 2005, \mnras, 359, L37 

\bibitem[Magnelli et 
al.(2013)]{mag13} Magnelli, B., Popesso, P., Berta, S., et al.\ 2013, \aap, 553, A132 

\bibitem[Mattila(2006)]{mat06} Mattila, K.\ 2006, \mnras, 
372, 1253 

\bibitem[Matsumoto et al.(2015)]{mat15} Matsumoto, T., Kim, 
M.~G., Pyo, J., \& Tsumura, K.\ 2015, \apj, 807, 57 

\bibitem[Matsumoto et al.(2011)]{mat11} Matsumoto, T., Seo, 
H.~J., Jeong, W.-S., et al.\ 2011, \apj, 742, 124 

\bibitem[Matsumoto et al.(2005)]{mat05} Matsumoto, T., 
Matsuura, S., Murakami, H., et al.\ 2005, \apj, 626, 31 

\bibitem[Matsuoka et al.(2011)]{pioneer} Matsuoka, Y., Ienaka, 
N., Kawara, K., \& Oyabu, S.\ 2011, \apj, 736, 119 

\bibitem[Mazin \& Raue(2007)]{max07} Mazin, D., \& Raue, M.\ 2007, \aap, 471, 439 

\bibitem[McCracken et 
al.(2012)]{mcc12} McCracken, H.~J., Milvang-Jensen, B., Dunlop, J., et al.\ 2012, \aap, 544, A156 

\bibitem[McVittie 
\& Wyatt(1959)]{mcv59} McVittie, G.~C., \& Wyatt, S.~P.\ 1959, \apj, 130, 1 

\bibitem[Mihos et al.(2005)]{mih05} Mihos, J.~C., Harding, 
P., Feldmeier, J., \& Morrison, H.\ 2005, \apjl, 631, L41 

\bibitem[Montes 
\& Trujillo(2014)]{mon14} Montes, M., \& Trujillo, I.\ 2014, \apj, 794, 137 

\bibitem[Nagamine et al.(2006)]{nag06} Nagamine, K., Ostriker, J.~P., Fukugita, M., \& Cen, R.\ 2006, \apj, 653, 881 

\bibitem[Negrello et al.(2010)]{neg10} Negrello, M., Hopwood, R., De Zotti, G., et al.\ 2010, Science, 330, 800 

\bibitem[Oliver et al.(2012)]{oli12} Oliver, S.~J., Bock, J., 
Altieri, B., et al.\ 2012, \mnras, 424, 1614 

\bibitem[Papovich et al.(2004)]{pap04} Papovich, C., Dole, 
H., Egami, E., et al.\ 2004, \apjs, 154, 70 

\bibitem[Partridge 
\& Peebles(1967a)]{par67a} Partridge, R.~B., \& Peebles, P.~J.~E.\ 1967a, \apj, 148, 377 

\bibitem[Partridge 
\& Peebles(1967b)]{par67b} Partridge, R.~B., \& Peebles, P.~J.~E.\ 1967b, \apj, 147, 868 

\bibitem[Presotto et 
al.(2014)]{pre14} Presotto, V., Girardi, M., Nonino, M., et al.\ 2014, \aap, 565, A126 

\bibitem[Puget et 
al.(1996)]{pug96} Puget, J.-L., Abergel, A., Bernard, J.-P., et al.\ 1996, \aap, 308, L5 

\bibitem[R Core Team (2015)]{r15} R Core Team.\ 2015, https://www.R-project.org/

\bibitem[Rafelski et al.(2015)]{raf15} Rafelski, M., Teplitz, 
H.~I., Gardner, J.~P., et al.\ 2015, \aj, 150, 31 

\bibitem[Rudick et al.(2011)]{rud11} Rudick, C.~S., Mihos, 
J.~C., \& McBride, C.~K.\ 2011, \apj, 732, 48 

\bibitem[Sanders et al.(2007)]{san07} Sanders, D.~B., 
Salvato, M., Aussel, H., et al.\ 2007, \apjs, 172, 86 

\bibitem[Scoville et al.(2007)]{sco07} Scoville, N., Aussel, 
H., Brusa, M., et al.\ 2007, \apjs, 172, 1 

\bibitem[Shanks et al.(1991)]{sha91} Shanks, T., 
Georgantopoulos, I., Stewart, G.~C., et al.\ 1991, \nat, 353, 315 

\bibitem[Smith et al.(2012)]{smi12} Smith, A.~J., Wang, L., 
Oliver, S.~J., et al.\ 2012, \mnras, 419, 377 

\bibitem[Somerville et al.(2012)]{som12} Somerville, R.~S., 
Gilmore, R.~C., Primack, J.~R., 
\& Dom\'inguez, A.\ 2012, \mnras, 423, 1992 

\bibitem[Stecker et al.(2006)]{ste06} Stecker, F.~W., Malkan, M.~A., \& Scully, S.~T.\ 2006, \apj, 648, 774 

\bibitem[Taniguchi et al.(2007)]{tan07} Taniguchi, Y., 
Scoville, N., Murayama, T., et al.\ 2007, \apjs, 172, 9 

\bibitem[Teplitz et al.(2013)]{tep13} Teplitz, H.~I., 
Rafelski, M., Kurczynski, P., et al.\ 2013, \aj, 146, 159 

\bibitem[Totani et al.(2001)]{tot01} Totani, T., Yoshii, Y., 
Iwamuro, F., Maihara, T., \& Motohara, K.\ 2001, \apjl, 550, L137 

\bibitem[Valiante et al. (2016)]{val16} Valiante E., et al.\ 2016, \mnras, submitted

\bibitem[van Dokkum et al.(2015)]{van15} van Dokkum, P.~G., 
Abraham, R., Merritt, A., et al.\ 2015, \apjl, 798, L45 

\bibitem[Viero et al.(2013)]{vie13} Viero, M.~P., Wang, L., 
Zemcov, M., et al.\ 2013, \apj, 772, 77 

\bibitem[Voyer et al.(2011)]{voy11} Voyer, E.~N., Gardner, 
J.~P., Teplitz, H.~I., Siana, B.~D., 
\& de Mello, D.~F.\ 2011, \apj, 736, 80 

\bibitem[Wang et al.(2014)]{wan14} Wang, L., Viero, M., 
Clarke, C., et al.\ 2014, \mnras, 444, 2870 

\bibitem[Wardlow et al.(2013)]{war13} Wardlow, J.~L., Cooray, 
A., De Bernardis, F., et al.\ 2013, \apj, 762, 59 

\bibitem[Watkins et al.(2014)]{wat14} Watkins, A.~E., Mihos, 
J.~C., Harding, P., \& Feldmeier, J.~J.\ 2014, \apj, 791, 38 

\bibitem[Wesson et al.(1987)]{wes87} Wesson, P.~S., Valle, 
K., \& Stabell, R.\ 1987, \apj, 317, 601 

\bibitem[Wesson(1991)]{wes91} Wesson, P.~S.\ 1991, \apj, 367, 
399 

\bibitem[Windhorst et al.(2011)]{win11} Windhorst, R.~A., 
Cohen, S.~H., Hathi, N.~P., et al.\ 2011, \apjs, 193, 27 

\bibitem[Windhorst et al.(2008)]{win08} Windhorst, R.~A., 
Hathi, N.~P., Cohen, S.~H., et al.\ 2008, Advances in Space Research, 41, 
1965 

\bibitem[Windhorst et 
al.(1984)]{win84} Windhorst, R.~A., van Heerde, G.~M., \& Katgert, P.\ 1984, \aaps, 58, 1 

\bibitem[Wright(2004)]{wri04} Wright, E.~L.\ 2004, \nar, 48, 465 

\bibitem[Wright et al.(2016)]{wri16} Wright, A., et al.\ 2016, \mnras, in press

\bibitem[Xu et al.(2005)]{xu05} Xu, C.~K., Donas, J., 
Arnouts, S., et al.\ 2005, \apjl, 619, L11 

\bibitem[Zamojski et al.(2007)]{zam07} Zamojski, M.~A., 
Schiminovich, D., Rich, R.~M., et al.\ 2007, \apjs, 172, 468 

\bibitem[Zemcov et al.(2014)]{zem14} Zemcov, M., Smidt, J., 
Arai, T., et al.\ 2014, Science, 346, 732 
 
\bibitem[Zwicky(1951)]{zwi51} Zwicky, F.\ 1951, \pasp, 63, 61 

\end{thebibliography}
\end{document}